\documentclass[a4paper,fleqn,usenatbib]{mnras}

\usepackage[T1]{fontenc}
\usepackage{ae,aecompl}

\usepackage{graphicx}	
\usepackage{amsmath}	
\usepackage{amssymb}	
\usepackage{mathtools}

\newcommand{\mydate}{Accepted by MNRAS, 05 July 2019; 
   in original form, 21 May 2019} 

\newcommand{\kms} {{\rm \, km \, s^{-1} }} 
\newcommand{\GAIA} {{\em GAIA}\ }
   
\newcommand{\kau} {\, {\rm kAU}} 
\newcommand{\msun} {\,M_\odot} 
 
\newcommand{\pc} {\, {\rm pc}} 
\newcommand{\kpc} {\, {\rm kpc}} 
\newcommand{\msecsq}{\, {\rm m \, s^{-2}}}

\newcommand{\vp} {v_{\rm p}} 
\newcommand{\vc}{v_{\rm C}} 
\newcommand{\vinf}{v_\infty} 
\newcommand{\rp} {r_{\rm p}} 
\newcommand{\mas} {\, {\rm mas}} 
\newcommand{\vratio}{\vp / \vc(\rp)} 
\newcommand{\newtwo}{  }  

\title[Gravity Test with 
 Wide Binaries in GAIA DR2]{Testing Modified Gravity 
 with Wide Binaries in GAIA DR2}
 
\author[C. Pittordis \& W. Sutherland]{
Charalambos Pittordis,$^{1}$\thanks{E-mail: c.pittordis@qmul.ac.uk}
Will Sutherland,$^{1}$\thanks{E-mail: w.j.sutherland@qmul.ac.uk}
\\
$^{1}$ School of Physics \& Astronomy, 
 Queen Mary University of London, Mile End Road, London E1 4NS, UK.\\
}

\date{ \mydate }

\pubyear{2019}

\begin{document}
\label{firstpage}
\pagerange{\pageref{firstpage}--\pageref{lastpage}}
\maketitle 

\begin{abstract}
 Several recent studies have shown that very wide binary stars
  can potentially provide an interesting test for modified-gravity
  theories which attempt to emulate dark matter; these systems should
  be almost Newtonian according to standard dark-matter theories, 
  while the predictions for MOND-like theories are distinctly
  different, if the various observational issues can be overcome. 
  Here we explore an observational application of the test from 
 the recent GAIA DR2 data release:
  we select a large sample of $\sim 24,000$ candidate wide binary stars
  with distance $< 200 \pc$ and magnitudes $G < 16$ 
  from GAIA DR2, and estimated component masses using a main-sequence 
 mass-luminosity relation.  
   We then compare the frequency distribution of 
   pairwise relative projected velocity (relative to circular-orbit
  value) as a function of projected separation; these distributions
  show a clear peak at a value close to Newtonian expectations, 
  along with a long ``tail'' which extends to much larger
  velocity ratios; the ``tail'' is considerably more numerous 
  than in control samples constructed from DR2 with randomised positions, 
  so its origin is unclear.  
 Comparing the velocity histograms with simulated data,  
  we conclude that MOND-like theories {\em without} an external field
  effect are strongly inconsistent with the observed data since
 they predict a peak-shift in clear disagreement with the data;  
  testing MOND-like theories {\em with} an external field effect 
  is not decisive at present, but has good prospects to become  
  decisive in future with improved modelling or understanding of
  the high-velocity tail, and additional spectroscopic data.  

\end{abstract}




\section{Introduction}
\label{sec:intro} 

Einstein's theory of General Relativity (GR) provides the best 
 known description of gravity on all scales. However,  
 much cosmological data (e.g. \citealt{Planck 2015}) 
 requires an additional cold, 
 non-baryonic \& non-visible dark matter
 (DM) component to match many observations, in addition to dark energy
  such as a cosmological constant.  
 At the present time there is no decisive direct detection of DM; 
 this leaves an open window for various modified-gravity theories,  
  which might potentially account for these various observations 
  without the requirement for exotic DM. 

The MOdified Newtonian Dynamics (MOND) is a well-known theory that attempts to
 explain weak-field/non-relativistic gravitational effects without DM. This
 theory was first proposed by \citet{Milgrom 1983} to explain the flat
 rotation curves observed in most spiral galaxies without requiring DM. 
 The original MOND formulation was non-relativistic and really a
  fitting function rather than a realistic theory; 
 it has later been incorporated into relativistic theories following from
 the Tensor-Vector-Scalar (TeVeS) theory proposed by
\citet{Bekenstein 2004}. While the original TeVeS is now excluded by
  the constraints on time-delay in the neutron-star merger
   GW 170817 \citep{gw-delay},  other versions remain viable; 
  see e.g. \citet{Clifton 2012} and \citet{Famaey 2012} for reviews
  of modified gravity, and e.g. \citet{Arraut14} for a non-local gravity
 model or \citet{Capozziello19} for recent cosmological comparisons.     

 Clearly, a convincing direct detection of dark matter would 
 be the most decisive scenario, but the converse is not true: null results
 from dark matter experiments can never rule out the paradigm,
 because the DM interaction cross-section might simply be too small 
  for any practical experiment (or, the cross-section could be 
  weak-like but the DM particle masses could be $\ga 10^9\, {\rm GeV}$, 
 implying a local number density far below the value for 
  conventional TeV-scale WIMPs). 
  Therefore, in the absence of a DM direct detection, new tests which can
  discriminate between DM and modified-gravity from direct
   tests of gravity at the relevant very low accelerations 
  are highly desirable.   
 
 Wide-binary stars (separations $\ga 5 \kau$) 
    are a promising route to a direct test,  
  since they have low orbital accelerations near or below the typical
  MOND acceleration scale $a_0 \sim 1.2 \times 10^{-10} \msecsq$, 
  and are generally presumed to contain negligible dark matter.  
 Studies of wide binaries in general have been explored 
  by e.g. \citet{Yoo 2003}, \citet{Lepine 2007}, \citet{Kouwenhoven 2010}, 
   \citet{jiang-tremaine}, \citet{slowpokes}, \citet{Coronado 2015}, 
   and others. 
 Previous work concerning tests of MOND-like gravity has been done by 
 \citet{Hernandez 2011},\citet{Hernandez 2012},
  \citet{Hernandez 2014}, \citet{Matvienko 2015}, \citet{Scarpa 2017} 
  and \citet{Hernandez 2019};    these typically give 
  hints of deviations in the direction expected from MOND-like gravity,
  though due to the limited precision of then-available data,
   these hints are not yet decisive.   

 In a recent paper \citep{PS18}, hereafter PS18, 
   we used simulations to explore the
  prospects for this test in anticipation of the much improved data expected 
  from the {\GAIA} spacecraft \citep{gaia-miss};  PS18 used 
   simulated wide-binary orbits 
  for a variety of acceleration laws, including Newtonian and various
  MOND models both with and without an external field effect (hereafter ExFE); 
 the general conclusion was that GAIA data provides promising 
 prospects for such a test, since MOND-like models {\em without} an ExFE
 should produce large and obvious deviations towards
  larger velocity differences. 

 In MOND-like models {\em with} the ExFE included, 
   as theoretically preferred,   the local Galactic
 acceleration substantially suppresses MOND-like effects, 
  but does not eliminate them. 
  These models give predicted relative velocities 
  much closer to Newtonian, but do still show
  subtle deviations, most notably a significantly larger fraction 
  of binaries with pairwise velocities in the  
  range $(1.1 - 1.5) \times v_c(r_p)$, where $v_c(\rp)$ is the 
  Newtonian circular velocity at projected separation $\rp$.  

 Here we recall two of the main conclusions from PS18: 
   MOND-like theories can allow bound binaries with relative
  velocities above the Newtonian ceiling, $v_{3D} / v_c(r_p) > \sqrt{2}$, 
  (where $v_{3D}$ is the 3-D pairwise relative velocity); but  
  with the ExFE included the fraction of such systems 
   is predicted to be very small, typically 1 percent or less;   
   so simply counting such systems is unlikely to be a practical test
   due to possible contamination, observational errors, and small-number
   statistics. 
 However, the upper percentiles of this velocity ratio, or similarly 
   the fraction of binaries with velocity ratio between
   $\sim 1.2 - 1.5$ are more promising statistics. 
 In PS18 we noted that no more than $11.1$ percent of 
  binaries should have a ratio $v_{3D} / v_c(r) > 1.2$ 
  in Newtonian gravity,  for {\em any} eccentricity distribution, 
  while plausible smooth eccentricity distributions produce
   a slightly lower percentage;     
  while MOND-like theories can produce a significantly higher
  percentage. 
   Since the 3D separation $r$ is not a practical observable 
   (since the line-of-sight separation is 
  typically well below the precision of distance measurements)  
   we have to replace it with projected separation 
    $r_p$ as a proxy, which shifts the ratios to lower values 
   depending on viewing angles; but this
   effect can be readily included in simulations. 
  If future observations were to measure a
   high-velocity tail of binaries well above the Newtonian 
   prediction (after statistical subtraction
  of contaminants),  
  this could in principle provide strong evidence in 
  favour of modified gravity.  

 Note in this paper, since radial velocities are not yet available  
  for the large majority of our binary candidates below, 
  we use 2D sky-projected 
  velocity differences rather than 3D velocities as in PS18; this shifts
   the most relevant velocity window downward to $\sim 1.1 - 1.5$, 
   and also implies
  that larger samples will be required to counter the added statistical 
   scatter from random viewing angles. 
  However, this does not substantially change 
   the general principle of the test; see Section~\ref{sec:3d2d}, 
  and also \citet{BZ18} and \citet{Banik 2019} for further discussion. 
 
 The second main conclusion from PS18 
  was that, in MOND theories {\em with} the ExFE,  
 there is an optimal window of projected separation, 
   $5 \la \rp \la 20 \kau$, for practical application of the test. 
   Even wider separations are not favoured in practice  
   because the inclusion of the ExFE causes the MOND-like effects 
    to almost saturate at $\rp \ga 10 \kau$, 
   while several observational issues become 
    proportionally worse at even wider separations.   

 The plan of this paper is as follows: in Section~\ref{sec:dr2} 
 we describe the selection of candidate wide-binaries from the
 GAIA DR2 data. In Section~\ref{sec:rand} we describe some randomised
  samples used to assess the number of chance projection systems,
  which turns out to be small.   
 In Section~\ref{sec:orbits} we discuss various simulations 
  of the velocity-ratio distributions 
  for Newtonian and MOND-like binary orbits (with and without the ExFE).
  In Section~\ref{sec:flyby} we discuss 
   simulations of velocity ratio for co-natal hyperbolic flyby systems. 
 In Section~\ref{sec:comp} we compare data and models, 
  finding reasonable agreement with a simulated ``Newtonian plus flybys''
    distribution; 
  and we summarise our conclusions in Section~\ref{sec:conc}.

\section{GAIA DR2 and sample selection} 
\label{sec:dr2} 

\subsection{Preliminary selection} 

Our starting point is the public GAIA Data Release 2 dataset 
  (hereafter DR2), 
 \citep{gdr2}, released on 2018 April 25. 
 We initially select all stars with measured parallax $\omega > 5 \mas$
 (i.e. estimated distance $< 200 \pc$)
  and GAIA broadband magnitude $G < 16$, yielding a sample of 970,760 stars. 
 (Data quality cuts are applied later on, in order that these
     may be adjusted post-selection). 
 The parallax and magnitude cuts above are chosen to provide a 
   large enough volume to contain a 
  usefully large statistical sample of wide binaries; while the moderate
   distance limit and relatively bright magnitude limit ensures that 
   GAIA provides high precision on distances and transverse velocities. 
  Finally,  the $G < 16$ cut ensures good feasibility for
 future follow-up high-resolution spectroscopy on 
   selected subsamples.  The sky distribution of these stars 
 shows a fairly uniform distribution, 
  with some enhancements near the Galactic Plane and some well-known open
  clusters. 


We then search this nearby-star sample for pairs of stars
 with projected separation $\le 40 \kau$ (calculated 
  at the mean distance of each candidate pair), 
  parallaxes of both stars consistent with each other within $4\times$
 the combined uncertainty,  and projected velocity difference $\le 3 \kms$ 
  as inferred from the difference in proper motions; here, the
   projected velocity difference is computed  
  assuming {\em both} stars in each candidate pair are
   actually at the mean of the two estimated distances. 
 
  We note here that this common-distance assumption is important:
   if the relative velocities are calculated using 
  individual parallax distances, then an example random 1 percent
  difference in parallax for a system
  with transverse velocity $40 \kms$ scales to a $0.4 \kms$ transverse 
   velocity difference, which is similar to or larger than the orbital
  velocities of interest below. 
   However, since we are almost entirely 
   interested in the velocity {\em difference} 
    within a binary, the common-distance assumption leads to an error
   in estimated relative velocity proportional to 
  the unknown {\em true} fractional distance difference, $(d_1 - d_2)/ d$ 
   (see also \citet{shaya-olling}, Section 2.4 of PS18 and 
 \citet{ElBadry19} for related effects).   
  For true binaries with random orientation we 
  expect $ \vert d_1 - d_2 \vert \le \rp$ for 71 percent of systems, 
  and $\le 2 \rp$ for 90 percent.  
  Then for a typical binary (see below) 
    with $\rp \sim 10 \kau$ and $d \sim 130 \pc$
   we have $\rp/d \sim 3.7 \times 10^{-4}$; this is 
    much smaller than the fractional uncertainty of the parallaxes,
   so choosing the common-distance assumption yields a much more 
   precise estimate of the relative velocity for {\em genuine} binaries. 

 This search results in a first-cut
 sample of $50,003$ candidate binaries, which is then pruned with
  additional cuts as described in the following subsections. 

\subsection{Sky cuts} 

Inspection of the initial sample showed a roughly uniform
 distribution across the sky,  
  with some enhancement near the
 galactic plane and around some well-known open clusters, 
 i.e. the Hyades, Praesepe and Upper Sco.  

 We therefore applied sky cuts to eliminate galactic latitudes
  $\vert b \vert \le 15 \deg$, and regions around the above clusters. 
 Our galactic-latitude cut removes about 1/4 of the sky, which is
   a moderate reduction in sample size but should significantly
 reduce confusion or contamination issues at high source densities. 
    These sky cuts reduce the sample to 33,667 candidate binaries.

\subsection{Triple and higher systems} 
\label{sec:triples} 
 To reject the majority of ``moving groups'' or similar,
 we searched our binary sample for any star in common between two 
 or more candidate binaries; if so, both or all those binaries 
  were rejected,  leaving a sample of 30,550 candidate binaries with no 
  star common to more than one candidate binary. 

 We also searched for additional co-moving companion stars 
  to a fainter limit: we selected
 a ``faint star'' sample of GAIA stars with $G \le 20$ and measured
 parallax $\omega \ge 4.2 \mas$;  for each star in a candidate binary,
 we then searched for faint-star companions with the following criteria: 
\begin{enumerate} 
 \item Parallax consistent with the main star at $4\sigma$. 
\item Angular separation less than $2/3$ of the main-binary separation 
  (since hierarchical triples are expected to be unstable for inner-orbit
  separation above $\sim 0.4 \times$ the outer separation); 
  and angular separation above $0.5$ arcsec to avoid barely-resolved
   companions.  
 \item Measured projected velocity difference from the main star 
    $\le 5 \kms$. 
\end{enumerate} 
 If any such ``third star'' was found, (in 375 cases), we rejected the 
 candidate binary since a hierarchical triple will generally
  boost the projected velocity difference of the wide pair; 
  this left a de-tripled sample of 30,175 candidate binaries.  

 Clearly, the third-star search above will not reject third stars
  which are either very faint or unresolved from one of our 
  binary members;   this will need to be considered for possible 
  followup observations later, but is unlikely to be the
   dominant source of contamination as we see below.

\subsection{Data quality cuts} 

We next applied data-quality cuts based on the GAIA 
 parameters, as \citet{Arenou18}  Equation 1 as follows: 
\begin{eqnarray} 
 \chi^2 & \equiv  & {\tt astrometric\_chi2\_al}  \nonumber \\
 \nu & \equiv & {\tt astrometric\_n\_good\_obs\_al } - 5 \nonumber \\
 u & \equiv & \sqrt{\chi^2 / \nu} \nonumber \\  
 u & \le & 1.2 \times {\rm max}(1, \exp\left[ -0.2(G-19.5) \right] ) 
\label{eq:arenou} 
\end{eqnarray} 

We rejected binaries where either star did not satisfy Eq.~\ref{eq:arenou}; 
this rejected another 5,270 candidate binaries, leaving a final cleaned sample
  of 24,282 candidate binaries which we use for the main analysis below. 

\begin{figure*}
\begin{center} 
\includegraphics[width=15cm]{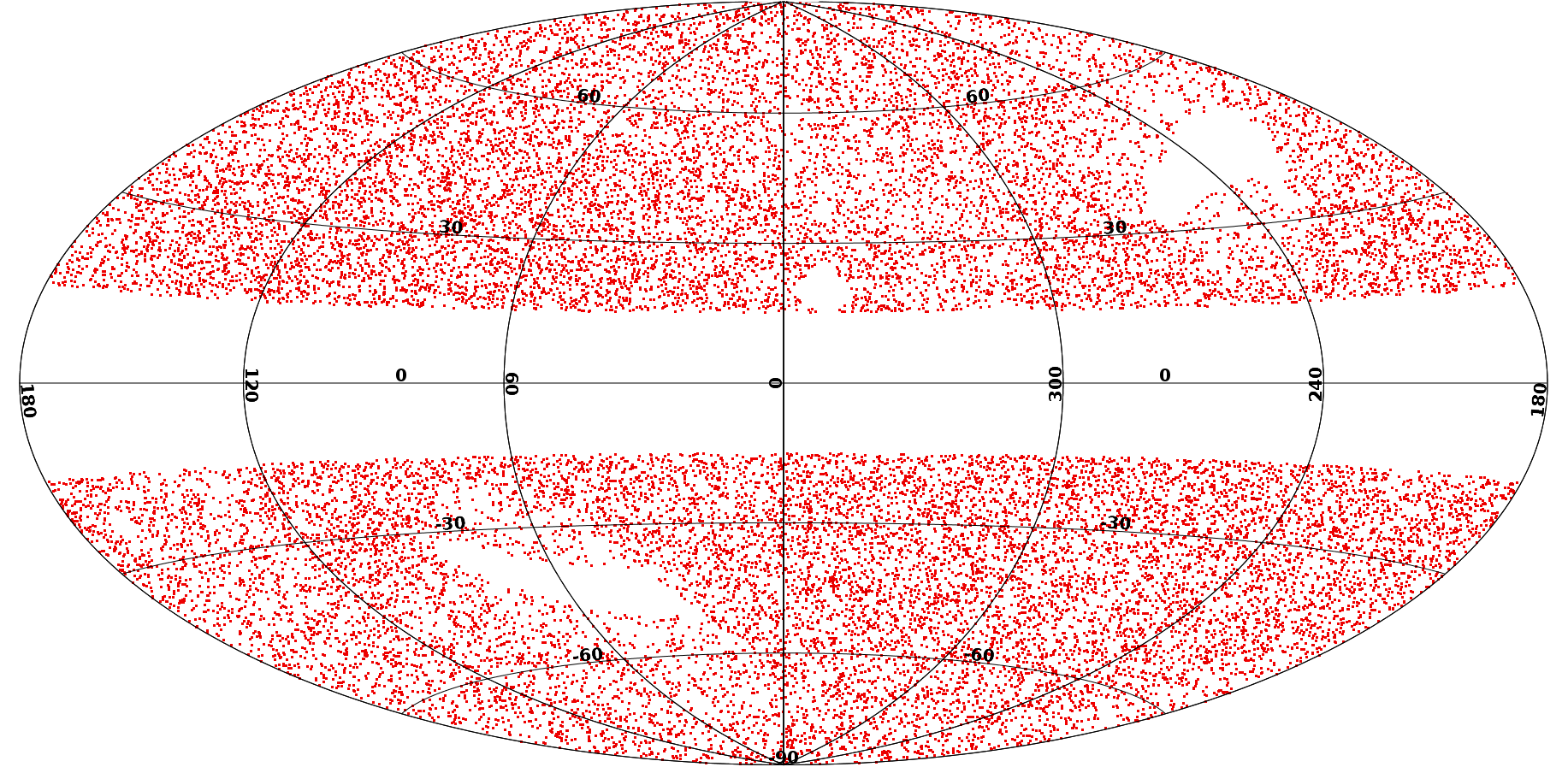} 
\caption{ 
 The sky distribution in Galactic coordinates for candidate
    binaries surviving all cuts in Section~\ref{sec:dr2}.  The two larger 
  holes are due to regions with fewer GAIA scans. } 
\label{fig:cutbins} 
\end{center} 
\end{figure*}

\subsection{Results and scaled velocities} 

 For the surviving 24,282 candidate binaries, we show
 a plot of projected velocity difference vs projected separation 
 in Figure~\ref{fig:rpvp}; this shows a clear excess approximately
 as expected for bound binaries, with an overdense cloud following
  a locus $\vp \sim 1 \kms (\rp / 1 \kau)^{-0.5}$. 
 We note that our sample starts to miss true binaries at 
   projected separations below  $\rp \la 0.6 \kau$, due to the
   $3 \kms$ velocity threshold, 
   but this $\rp$ is much  smaller than the separations of interest 
  below. At $\rp > 5 \kau$ the threshold includes pairs with velocity 
  difference far above the bound limit, 
  which are interesting for assessing sample contamination as seen below.

\begin{figure}
\begin{center} 
\includegraphics[width=9cm]{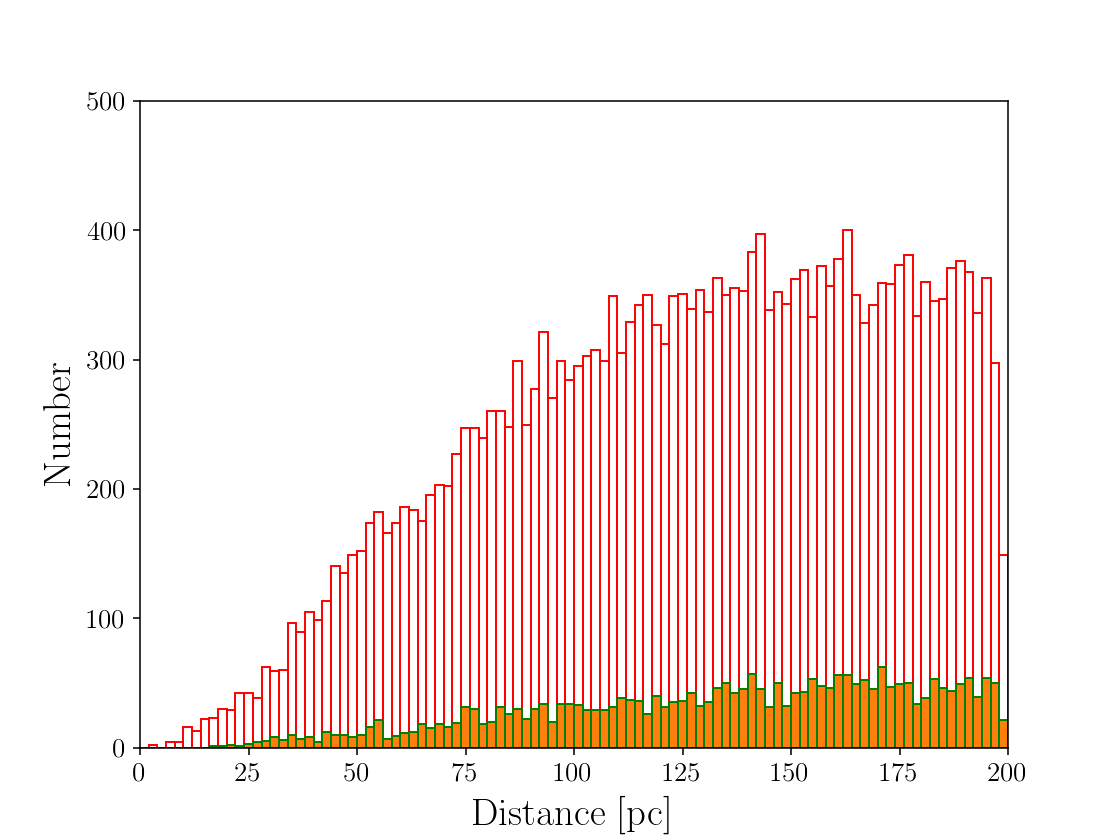} 
\caption{Histogram of average distance for the candidate binaries. 
  The open histogram shows all 24,282 binaries passing the cuts, 
 while the filled histogram shows the subset with $5 < \rp < 20 \kau$. }
\label{fig:dhist} 
\includegraphics[width=9cm]{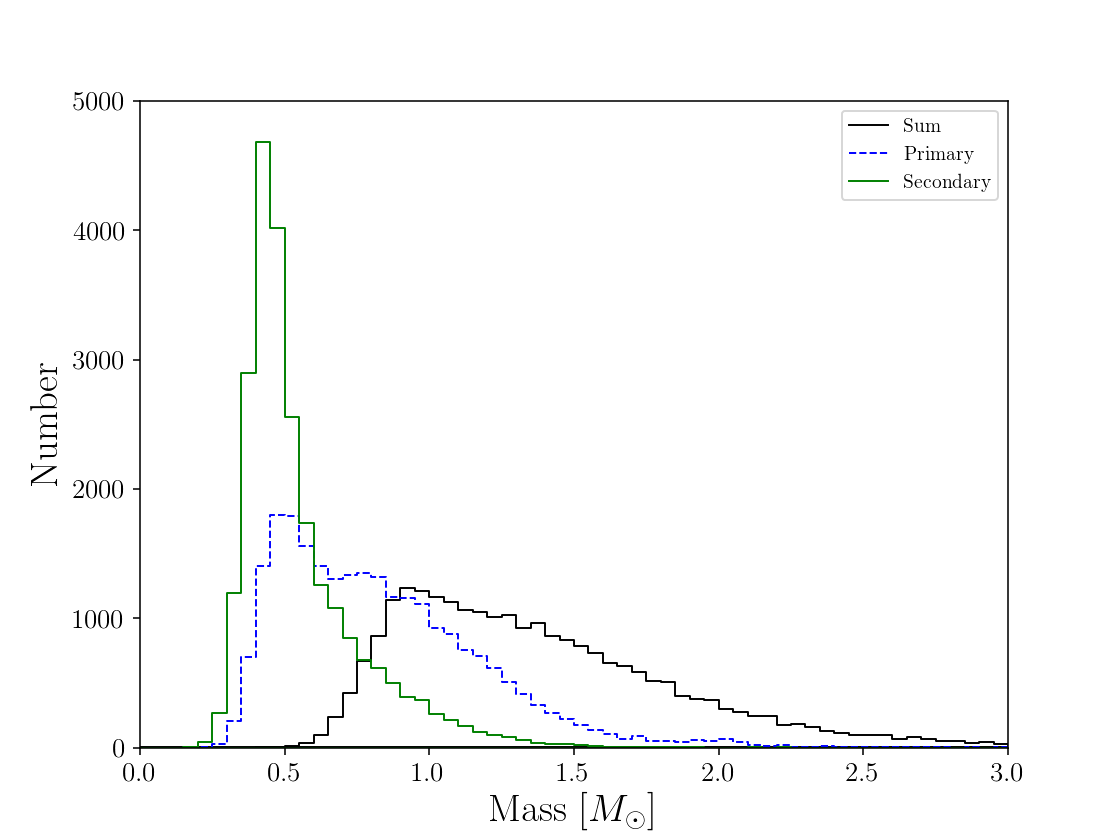} 
\caption{Histogram of estimated masses for the candidate binaries. 
  The black solid line shows combined system mass; dashed blue line
 shows the primary (more massive) star, and green line shows
 the secondary star. } 
\label{fig:mhist} 
\end{center}
\end{figure}

\begin{figure*} 
\begin{center} 
\includegraphics[width=16cm]{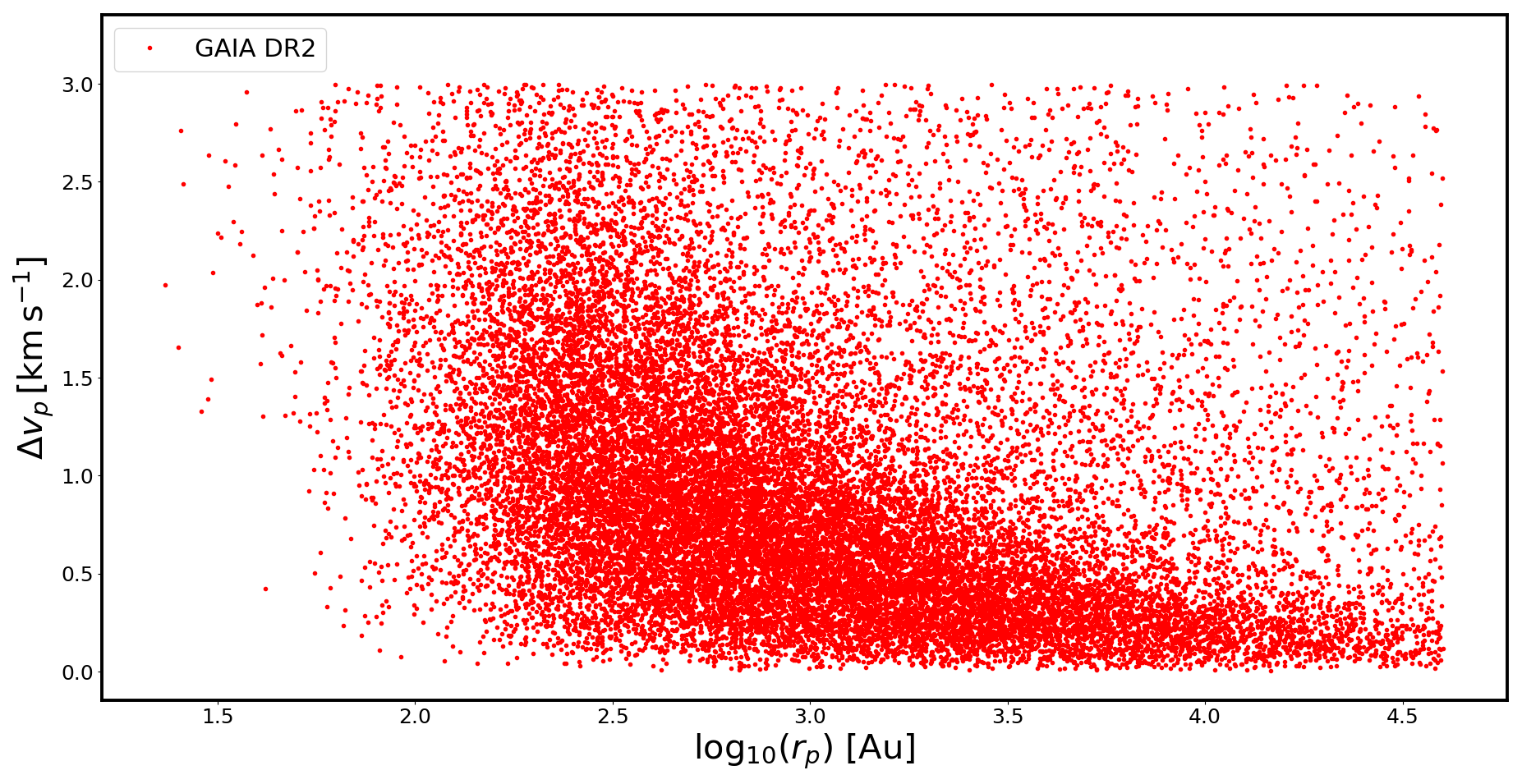} 
\caption{Scatter plot of projected velocity $\vp$ (y-axis)
 vs projected separation (log scale, x-axis) for the cleaned binary
 sample.  The main selection cuts are visible at top and right. } 
\label{fig:rpvp} 

\includegraphics[width=16cm]{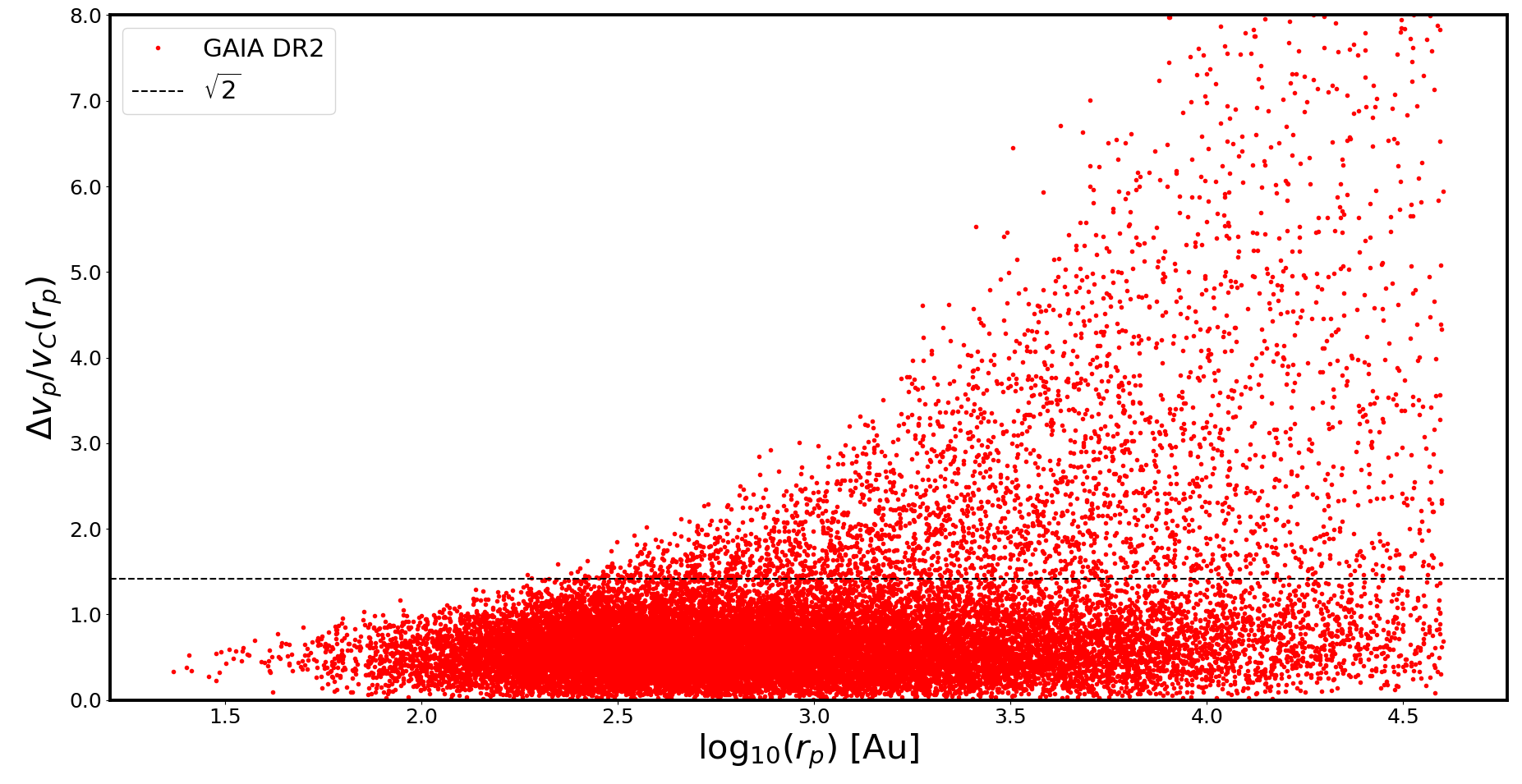} 
\caption{Scatter plot of projected velocity relative to Newtonian, 
 $\vp/v_c(\rp)$, vs projected separation.  The dashed line at 
 $\sqrt{2}$ indicates the Newtonian limit. The upper cutoff is 
   now slightly fuzzy due to the additional dependence on mass. } 
\label{fig:rp_vratio}
\end{center}
\end{figure*}

 It is more informative to rescale to the typical Newtonian
  orbit velocity, so we next estimate masses for each binary 
  using an estimated mass/luminosity relation:  here,
 we adopt the main-sequence $M_I(mass)$ relation from \citet{PecMam2013}, 
 and the $V - I, M_I$ colour relation from the same,
  where $M_I$ denotes absolute magnitude.  From those we
  apply the colour relation given in Table~A2 of \citet{Evans18} to predict
  $G$ magnitude from $V$ and $I$ magnitudes as 
 \begin{eqnarray} 
  G & \simeq & V \; - \; 0.01746 \; + \; 0.008092 (V-I) \; - \; 
     0.2810 (V-I)^2 \nonumber \\ 
  & &  + 0.03655 (V-I)^3 
 \label{eq:gvi} 
 \end{eqnarray} 
 to obtain 
  a predicted relationship between absolute GAIA magnitude 
   $M_G$ vs mass; we then fit to this to obtain an approximate
 mass/$M_G$ relation 
\begin{equation}
 \frac{M}{\msun} = 10^{0.0725 (4.76 - M_G )}  
\label{eq:mass-mg} 
\end{equation} 
 Then, for each star we have $M_G$ directly from $G$ and 
  parallax distance, 
  hence an estimated mass follows.   Since the luminosity(mass) 
   relation is rather steep, small errors in $G$ or distance
  have relatively little effect on mass estimates below.   
 A histogram of estimated distances for our binary sample is shown 
 in Figure~\ref{fig:dhist}, and a histogram of estimated masses 
 is shown in Figure~\ref{fig:mhist}. 

 For each candidate binary we then define 
\begin{equation} 
 v_c(\rp) \equiv \sqrt{G M_{tot} / r_p} 
\label{eq:vc} 
\end{equation} 
 as the estimated circular-orbit velocity at the current 
  {\em projected} separation; 
 for each candidate binary, we then divide the measured projected velocity 
  difference by the above to obtain a dimensionless ratio 
 $v_p / v_c(\rp)$; a scatter plot is shown in 
 Figure~\ref{fig:rp_vratio}, and 
 various histograms of this ratio  are compared with models below.

\subsection{Transverse velocity errors} 

\begin{figure}
\begin{center} 
\includegraphics[width=8cm]{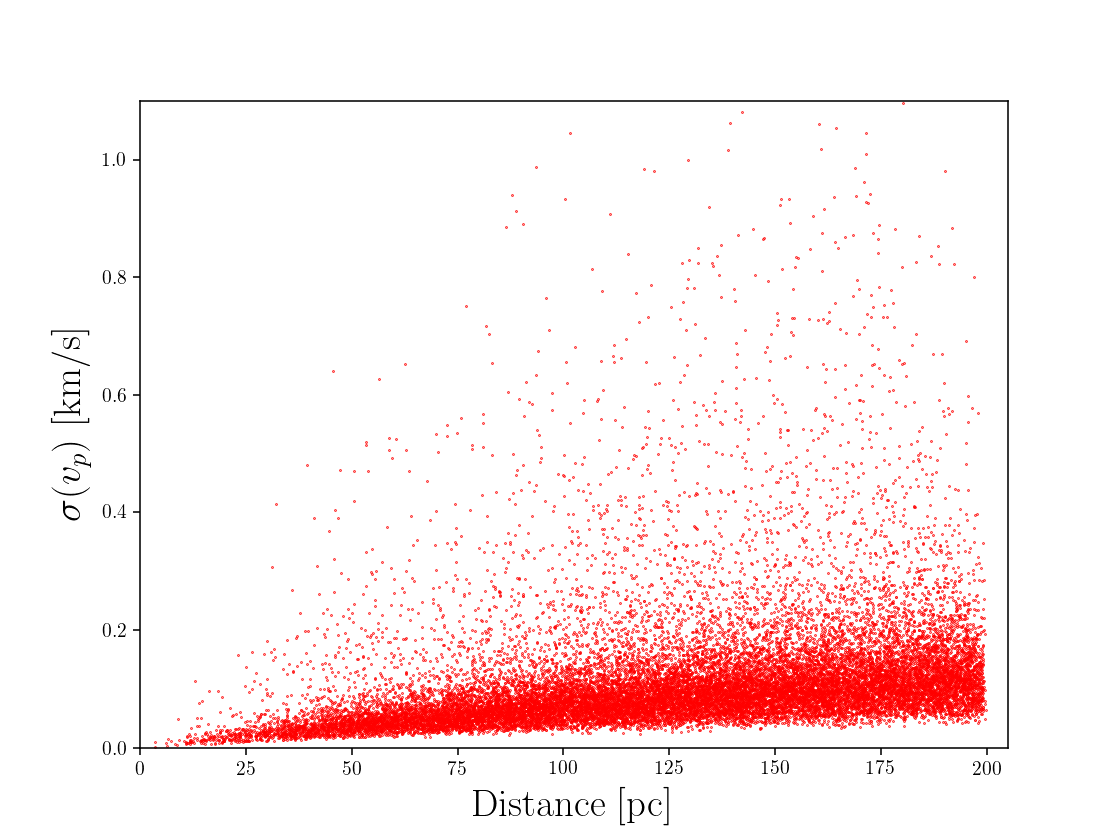} 
\caption{Scatter plot of rms velocity uncertainty $\sigma(\vp)$ versus
 mean distance for the 24,282 binaries surviving all cuts.}
\label{fig:dsigv} 
\end{center}
\end{figure}

 We have estimated relative-velocity errors assuming
 uncorrelated errors between the two components of the binary, 
  simply from the root-sum-square of the quoted rms errors in $\mu_\alpha$
 and $\mu_\delta$ for each of the two stars in each binary, 
  and multiplying by distance to obtain the transverse-velocity 
 error. (This should be reasonable as long-range correlated errors should 
  mostly cancel between the two stars).  
  The  median of this for the binary sample above is 
  $\sigma(\vp) \approx 0.09 \kms$, which
  is already impressively small.  A scatter plot of $\sigma(\vp)$
 versus distance is shown in Figure~\ref{fig:dsigv}; the trend
 with distance is clear, but
 most systems have $\sigma(\vp) \la 0.15 \kms$ even near our
 $200 \pc$ limit.

Converting to the ratio to circular-orbit velocity, 
  $\sigma(\vp)/v_c(\rp)$, the median for the
 full candidate sample is 0.08 and the 80th percentile is 0.14; 
  for the ``wide'' subsample with $5 < \rp < 20 \kau$, the
  median is 0.23 and the 80th percentile is 0.39. 

The latter values are significantly smaller than 1, but not very small, 
  so the effect of random proper motion errors will affect the detailed
 shape of the distributions below.  However, in future GAIA
 data releases these values are expected to reduce by factors
 of at least 2--4 as  proper motion precision scales
  as $\propto t^{-3/2}$,  so the random errors in proper motions  
  are likely to become relatively unimportant in the medium-term 
  future. 

  We note that for a ``typical'' binary below 
  at $\rp \sim 10 \kau$ and $d \sim 130 \pc$,
   the angular separation is $0.37$ mrad
  or 77 arcsec, so these are very well resolved and the uncertainty
   on $\rp$ is essentially the same as the error on the mean distance,
   typically well below 1 percent and almost negligible.  
  The error on $\vp$ is dominated 
   by random errors on the proper motions, assuming that
  correlated systematic errors mostly cancel between the two components
  of the binary.   Since we are mostly interested in statistical
  distributions, the effect of random errors is modest as long
   as these are not larger than $\sim 0.25$ in $v_p / v_c(\rp)$.  
  Note that for systems with small observed ratios 
  $\vp/v_c(\rp) \sim 0.5$, the {\em fractional} uncertainty in this
  ratio is rather large; 
   however such systems still have a high probability
  of the true ratio being $\la 0.8$, so this scatter is relatively
   unimportant.   For systems with $\vp/v_c(r_p) \ga 1$, the fractional
 uncertainty is relatively modest; though possible non-Gaussian
  errors in the GAIA data remain a concern, this should improve in 
  future GAIA releases as more observing epochs become available
   to reject outliers.

\subsection{Comparison with El-Badry \& Rix} 

 Here we note that a catalogue of candidate binaries 
 in GAIA DR2 has been published by \citet{ElBadry18} (hereafter ER18). 
 Much of our sample selection was completed independently 
  prior to the appearance of ER18, but here we give a brief comparison. 

 ER18 chose (coincidentally) the same $200 \pc$ limit as here, 
 but there are two  main differences: firstly,  our selection adopts a fixed
  threshold $v_p \le 3 \kms$, whereas ER18 use a separation-dependent 
   threshold  which translates to $v_p \le 2.1 \kms (\rp / 1\kau)^{-0.5}$ 
  (equivalent to the Newtonian bound limit $\sqrt{2} \vc$ for 
   a $2.5 \msun$ system, or just above at 
  $1.83 \vc$ for a more typical system with mass  
   $1.5 \msun$).  
  This means that our sample extends to substantially higher
  (unbound) velocity ratios at $\rp \ga 3 \kau$, which turns out to be
   useful below for investigating the high-velocity tail.  
 Secondly, our sample uses a magnitude limit $G < 16$, while ER18 have 
  a $G < 20$ limit (but also other cuts on relative errors, 
  which do introduce  some implicit magnitude-dependence). 

 There are additional differences in how we cut for clusters, 
  triples, etc, but these turn out to be relatively less important. 
 
 We have done a cross-match of our sample to ER18 as follows: 
  selecting the subset of the ER18 sample where both stars have $G < 16$, 
   our galactic cut $\vert b \vert > 15 \deg$ and $\rp < 40 \kau$ gives
  a subsample of 18,513 candidate binaries from ER18 which could 
  in principle pass our other cuts.    Cross-matching those against
 our cleaned sample of 24,282 candidates, we find that 15,652
  are in common.   Most of the additional binaries in our sample
  are either at larger velocity differences (above the ER18 limit and below
  our $3 \kms$ limit), or at smaller separations 
   $\rp < 0.5 \kau$ which we do not consider below. 

  Considering a subsample of our candidates with
   $3 \le \rp \le 20 \kau$ and velocity ratio $\le 2$, we find 
   3380 candidates;  of those, 3138 are common to the ER18 sample, 
   while only 242 are not in ER18.  (There are an additional 606 candidates 
    in ER18 satisfying the above criteria,  but not in our cut sample above). 
  This indicates that the different selection criteria have not had
   a major influence in this region.  

\section{Random samples} 
\label{sec:rand}

\subsection{Construction of random samples} 

It is clearly important to estimate the level of contamination 
 of our sample by random chance projections of unrelated stars
 which just happen to have chance small velocity differences. 
 To do this, we have constructed several randomised samples by 
 first removing one star from each binary, randomising the 
 true RA/Dec values by a few degrees (see below) in each coordinate,
 then re-running the binary search on the position-randomised sample.
 
 Here, the removal of one star from each binary was chosen since
  otherwise close binaries can be ``scattered'' and re-selected
 as wide binaries in the randomised list.  Also, the choice of
 few degree position shifts is small enough to preserve the
  global distributions with respect to galactic coordinates, 
 but is large enough  to eliminate most truly associated stars.   

\begin{figure*} 
\begin{center} 
\includegraphics[width=16cm]{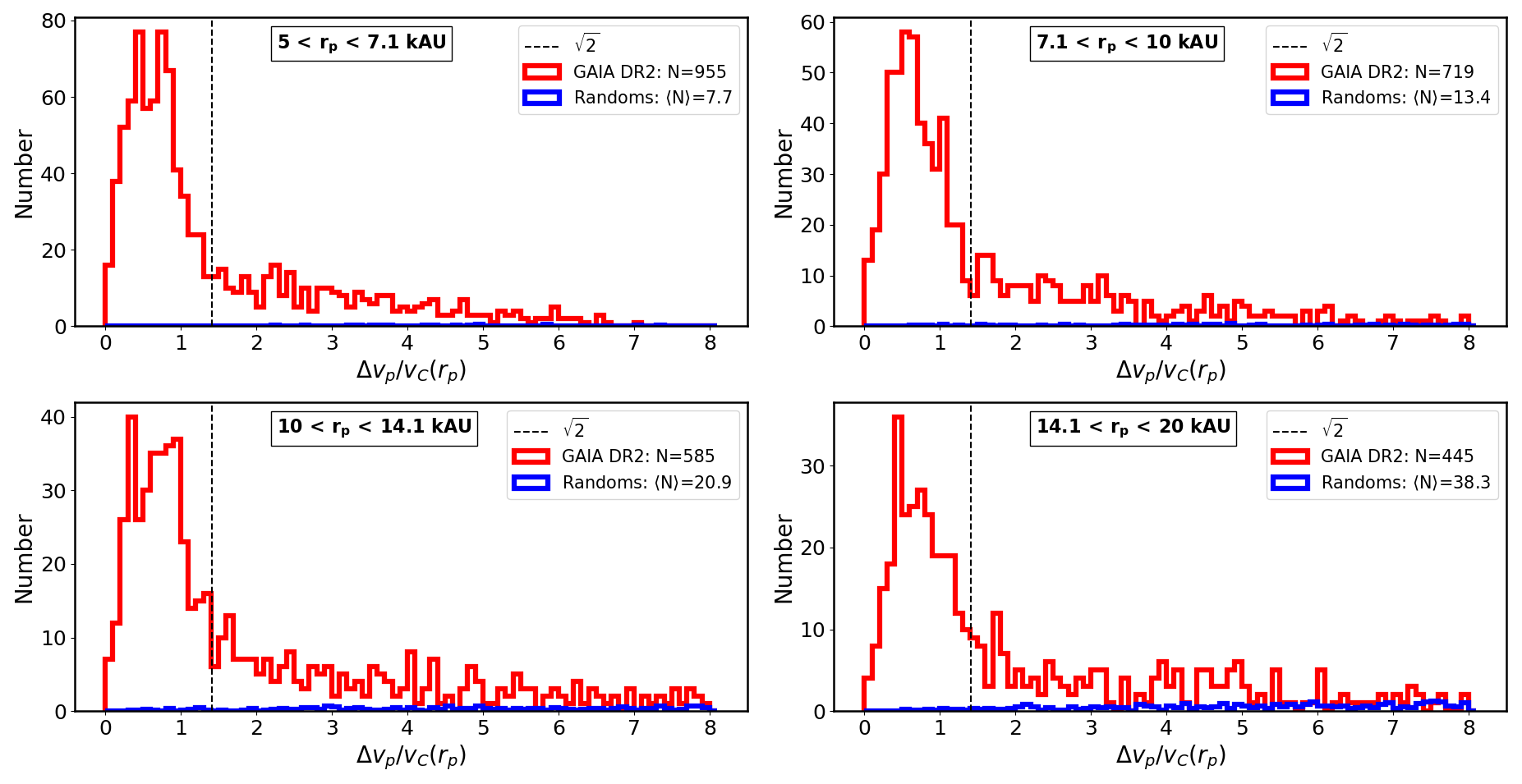} 
\caption{Histograms of velocity ratio $\vp/v_{c}(\rp)$ for binaries in GAIA
  data (red), and the mean of 9 randomised samples (blue). 
   The four panels show ranges of projected
 separation, $5 - 7.1 \kau$ up to $14.1 - 20 \kau$ as labelled.  
  Sample numbers are shown in the legend. } 
\label{fig:vrhist520} 
\includegraphics[width=16cm]{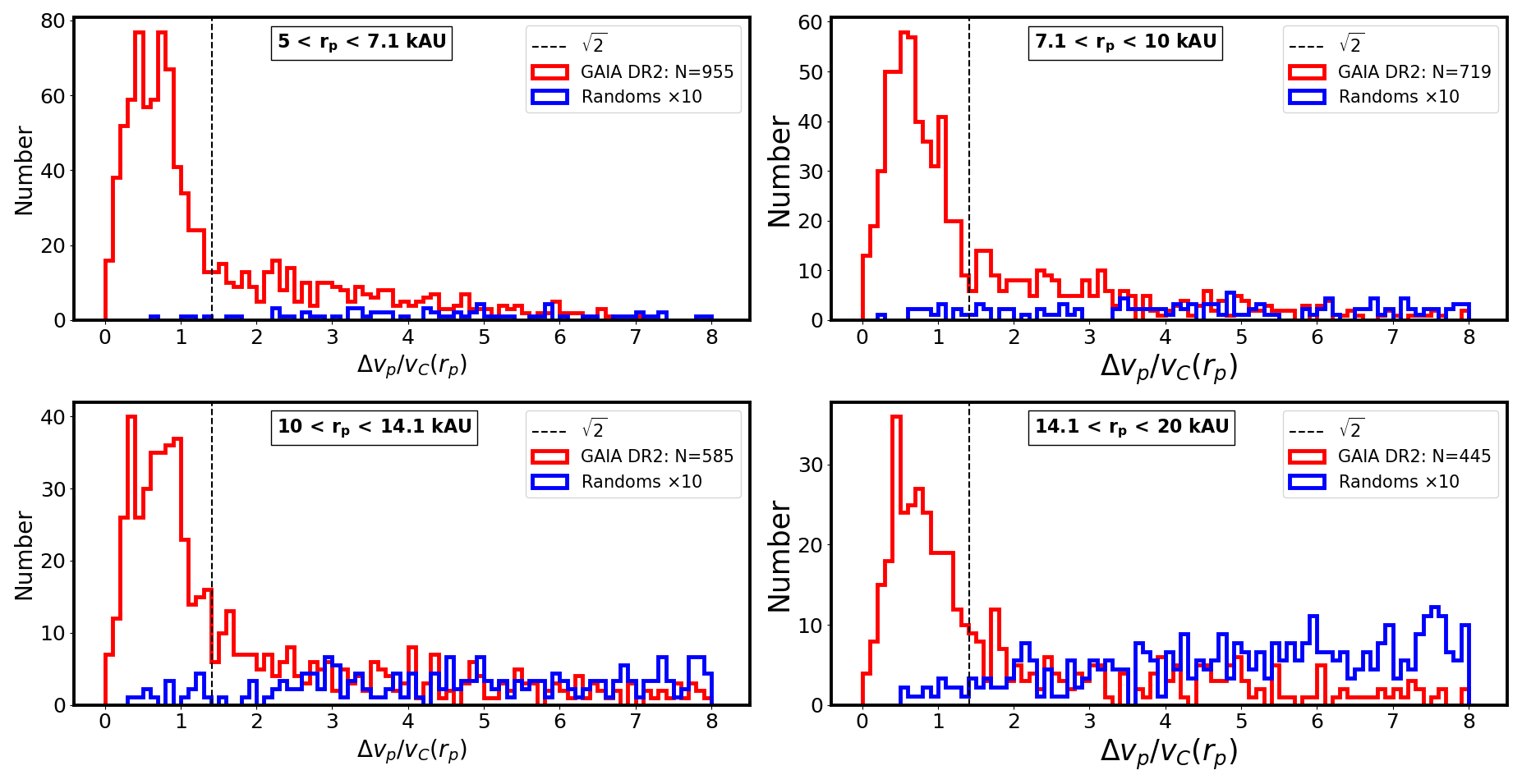} 
\caption{Same as Figure~\ref{fig:vrhist520},  but with the 
  mean random samples artificially scaled-up by $10\times$ 
  to enhance visibility.} 
\label{fig:vrhistx10} 
\end{center}
\end{figure*}

 We show an example histogram for a set of randomised samples
 compared to the data in Figure~\ref{fig:vrhist520}.  Since there
 the random counts are much smaller than the data and hard to see, 
  the following Figure~\ref{fig:vrhistx10} shows the same 
  with the random counts multiplied by $10\times$ for visibility.  
 
\subsection{Comparison of data and randomised sample} 

The main result from our randomised samples is that they contain
  far fewer candidate ``binaries'' than the actual GDR2 data: 
 the data samples outnumber the mean randoms by a factor of 
  124 in the separation bin 5 to 7.1 kAU,  
  falling to $11.6$ in our widest bin 14 to 20 kAU. 
 As expected the randomised samples show no peak at small velocity ratios, 
  but a fairly smooth distribution with a gradual rise towards
   larger velocity ratios.  

 We note that this large excess number in the real sample
   is present especially at ``bound'' velocity ratios $< \sqrt{2}$, 
  but also persists to large velocity ratios $\ga 2$. 

 It seems reasonable to assume that the clear peak in the 
  data at small velocity ratio $\sim 0.6$  is dominated
  by genuine bound binaries,  while the ``tail'' at 
  larger velocity ratios is rather unexpected: the tail is 
  much more populous than our randomised samples, so it
 is clearly not due to chance projections of unrelated stars.  
  But the tail extends smoothly to velocity ratios much larger 
  than attainable even in  reasonable modified-gravity models; 
 thus the tail may be
  dominated by stars with some common origin, e.g. co-natal pairs
  of stars born in the same open cluster which has subsequently
  dissolved, while the two star velocities retain a memory of their
  common origin, leading to unbound but correlated pairs in phase-space.  
 It seems clear that understanding and modelling the origin of this
 tail will be crucial to use the binaries for a gravity test: 
  this is modelled later in Section~\ref{sec:flyby}.   
 Before investigating the tail, 
  we next discuss simulations of velocity ratios 
 for genuine bound binaries. 

\section{Orbit simulations} 
\label{sec:orbits} 

\subsection{Velocity ratios} 
\label{sec:3d2d} 

We here recall that, unlike PS18 where we used
 3D relative velocities and 2D projected separations, 
 in this paper we are using 2D sky-projected velocities as
 well as 2D projected separations; this is simply because
 radial velocities are not yet available for the large majority
 of our candidate WBs, though they should become available in future
  with large spectroscopic surveys 
  such as 4MOST, PFS, WEAVE and MSE, and targeted followup.   

For the 2D velocity ratios as above, note that 
 the true value (in the absence of 
   perspective rotation effects and/or observational errors), 
  is always smaller than the corresponding (unknown) 
  value in 3D, since $\vp \le v_{3D}$, and $r_p \le r$ 
  so $v_c(r_p) \ge v_C(r)$, by a factor which depends
 on the unknown alignment and orientation of the orbit. 

\begin{figure}  
\begin{center} 
\includegraphics[width=8cm]{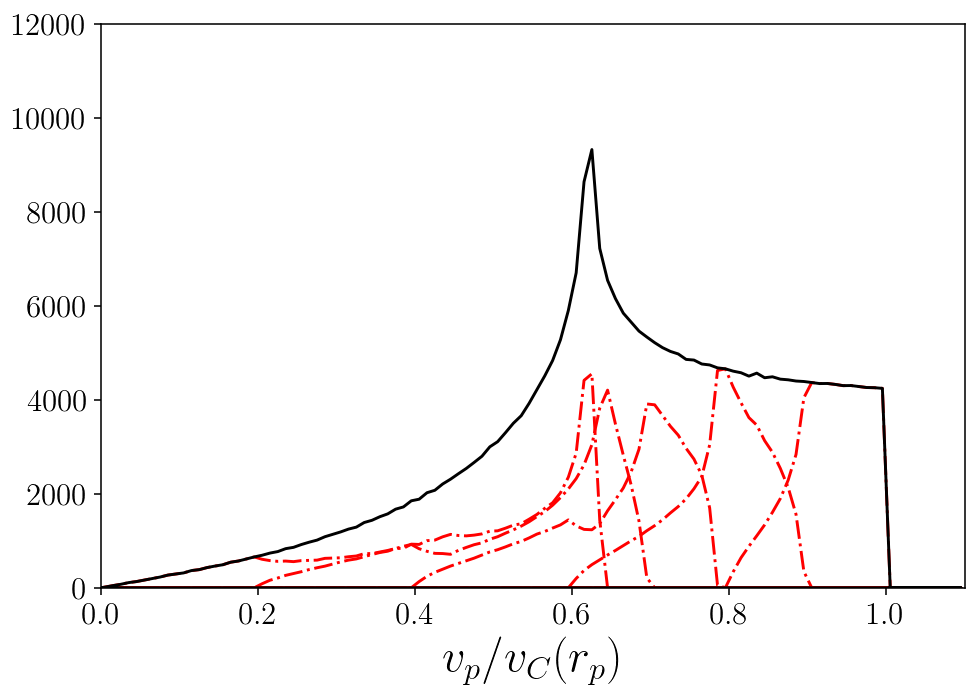} 
\caption{Distribution of 2D velocity ratio $\vp / v_c(\rp)$ for
 randomly-viewed circular orbits. 
  The black solid curve shows the full distribution,
 while red dash-dot curves show contributions from 5 equal bins of 
  $\vert \cos i \vert$,  from
 $0 < \vert \cos i \vert < 0.2$ (left) to $0.8 < \vert \cos i \vert \le 1$ 
 (right). } 
\label{fig:vratcirc} 
 \end{center}
 \end{figure}

 We note here that for circular orbits the 3D ratio is 1 by definition,
  and the above 2D ratio is readily derived as 
\[ 
 \vp / v_c(\rp) = \sqrt{1 - \sin^2 i \sin^2 \phi} 
  \left[ 1 - \sin^2 i \cos^2 \phi \right]^{1/4} 
\]
 where $i$ is inclination and $\phi$ is orbit angle from conjunction; 
  here the first factor is due to the deletion of the line-of-sight velocity
 component, while the second factor is $\sqrt{r_p/r}$ 
   due to projected separation. 
 A histogram of this ratio 
  for random angles is shown in Figure~\ref{fig:vratcirc}; 
   this is explained because nearly face-on orbits $\sin^2 i \la 0.36$ 
  produce a ratio close to 1 at all orbit phases,
  while nearly edge-on orbits $\sin^2 i \ga 0.8$ 
  produce a maximum value near 0.62 at intermediate angles 
   $\phi \sim 0 \pm 0.62$ or $\pi \pm 0.62$ rad 
   from conjunction, dropping to small values near 
  conjunction or greatest elongation;   
   thus high-$i$ orbits create the caustic spike and 
  the tail to low ratios, while low-$i$
  orbits fill the plateau from 0.7 to 1.   

 \begin{figure}  
 \begin{center} 
\includegraphics[width=8cm]{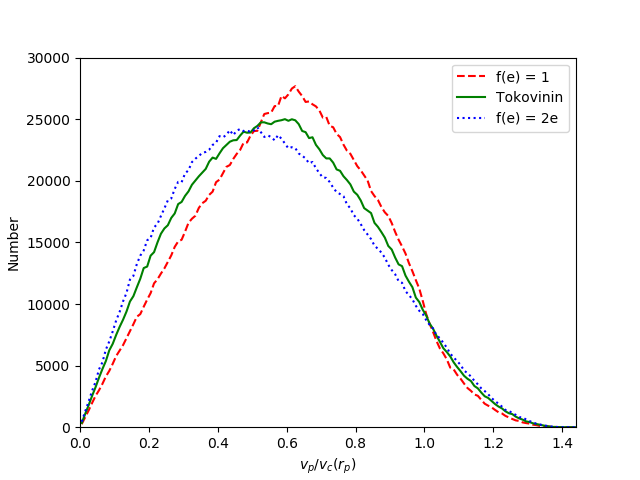} 
\caption{Distribution of 2D velocity ratio $\vp / v_c(\rp)$ for
 randomly-viewed elliptical orbits, with 3 example eccentricity distributions: 
 dashed red curve shows flat distribution $f(e) = 1$; 
 solid green curve shows Tokovinin distribution 
  $f(e) = 0.4 + 1.2e$; dotted blue curve shows $f(e) = 2e$. } 
\label{fig:vratell} 
\end{center}
\end{figure} 

 The above shows a simple but unrealistic case of pure circular orbits: 
  for realistic distributions of non-zero orbit eccentricity, we
  calculate the resulting distribution via numerical simulations, 
   similar to PS18 but using 2D projected velocity differences 
  instead of 3D, i.e. suppressing the radial-velocity information.   
 Some example results for Newtonian gravity and three selected eccentricity
  distributions are shown in Figure~\ref{fig:vratell}; 
 the eccentricity distributions are flat ($f(e) = 1$), 
   the observational fit 
  of \citet{Tokovinin 2016} which is $f(e) = 0.4 + 1.2e$; 
  and finally $f(e) = 2e$. 
   As expected the distributions become broader and 
  extend to values $> 1$, while the circular-orbit spike near 0.62 
  is smeared out, resulting in a broad peak at a ratio around 0.6 
  and a steeply-declining tail at values above 1.   


 A notable point here is that the tail is relatively small,
   since a ratio $\ga 1.1$ requires a simultaneous combination of three 
   factors: a fairly eccentric orbit, a viewing angle roughly face-on, 
   and a epoch fairly close to orbit pericenter today.  Each of these
  singly is only mildly improbable, but they are expected to be 
   almost independent of each other so the combination of all three 
    has a rather small probability. 
  Also, Figure~\ref{fig:vratell} shows
 that the tail at velocity ratio $\ge 1.1$ is 
 relatively insensitive to the details of the eccentricity
  distribution, similar to the PS18 result 
  for 3D velocities (see also Fig.~3 of BZ18 for a 
 similar conclusion); this is because extreme-$e$ orbits $e > 0.9$ 
  achieve higher maximum ratios, but spend less time near pericenter, 
  so much of the tail near velocity ratio $\sim 1.1$ is contributed by 
   moderate-$e$ orbits.  

  In particular, it is possible to get {\em fewer} binaries at high ratios
   if orbits are preferentially near-circular, but the converse is not
  true i.e. the Tokovinin or $f(e) = 2e$ distributions are almost a ceiling
   on the fraction of bound binaries in the high-velocity-ratio tail 
 \footnote{Note that it may be possible to get a ``false negative'' conclusion
  if a MOND-like gravity modification is correct, but there is also a strong
  bias against eccentric orbits at larger separations which cuts
  off the high-velocity tail.  This seems rather contrived, and may be 
  testable by looking at detailed shapes of the distributions.}.  
  Also, the predicted distribution 
 is steeply declining at ratios $\sim 1.0 - 1.3$; 
   this feature is useful later, since a rather moderate shift 
  in velocities from MOND-like gravity with the ExFE 
  (below, we find typically $\sim 15\%$ velocity boost 
     above Newtonian) 
  translates into a rather large multiplicative enhancement 
   (typically $2\times$ to $2.5\times$)  
    in the predicted fraction of true binaries with
    velocity ratio $\ga 1.1$; this large enhancement 
   cannot be mimicked in standard gravity  
   by varying the eccentricity distribution,  
   in agreement with BZ18.    

\subsection{Modified gravity orbits} 
\label{sec:orbsmg} 
 Here we simulate a large sample of $\sim 5 \times 10^6$ orbits
 with random values of $a, e$ in several modified-gravity theories 
  then study the joint distribution of observables, 
 in particular projected separation 
  $r_p$ and relative velocity ratio $\vp / v_c(r_p)$, as defined above.  
 Here we use the \citet{Tokovinin 2016} eccentricity distribution,
  as the intermediate of the 3 example cases above. 

 In the case of modified gravity models, the orbits are generally 
 not closed ellipses, so they are not strictly defined by 
 the standard Keplerian parameters $a, \, e$, but we still need
  to simulate a distribution in size and shape of orbits.  
  To deal with this, as in PS18, for a modified-gravity orbit 
   we define an ``effective'' orbit size $\hat{a}$ and quasi-eccentricity 
   $\hat{e}$ as follows: we define 
   $\hat{a}$ to be the separation at which the simulated relative velocity
   is equal to the circular-orbit velocity (in the current 
   modified-gravity model), then
   we define $\theta_{\rm circ}$ to be the angle between the relative velocity 
   vector and the tangential direction when the orbital separation crosses
    $\hat{a}$, 
  and then $\hat{e} \equiv \sin \theta_{\rm circ}$;  these definitions 
   coincide with the usual Keplerian $a, e$ 
  in the case of standard gravity. 


After integrating these orbits using one of a selected set
 of gravity laws (Newton/GR, MoND with/without ExFE) and a chosen
 value for external field $g_e$,  we ``observe'' the
 resulting binaries at many random times and random inclinations to the 
  line-of-sight.  

 For each simulated orbit/epoch snapshot, we produce simulated 
 observables including the projected separation $r_p$, projected relative
 velocity $\vp$, and also $\vp/v_C(r_p)$ corresponding
 to our observable from GAIA. 

The radial acceleration law is chosen according to the
 selected gravity theory under consideration, and also with the
 external field effect turned off or on (see below).  
 For the Newtonian/GR case, we have the standard 
\begin{equation}
g_{N} = \frac{G(M_1+M_2)}{r^2}
\label{eq:g_n}
\end{equation}

For the MOND-like case (both with and without the ExFE), 
 we use the fitting function of \citet{McGaugh 2016} (hereafter MLS), 
  sometimes known as the ``radial acceleration relation'', 
  given by 
\begin{eqnarray} 
 g_{MLS} & = & g_N \, \nu(g_N/a_0) \ ; \nonumber \\ 
 \nu(y) & = & \frac{1}{1 - \exp(-\sqrt{y})} \ ; 
\label{eq:g_mls} 
\end{eqnarray} 
 we refer to this $\nu$ function as the MLS interpolating function below. 
 This function is shown by MLS to produce a good fit to rotation curves
  for a large sample of disc galaxies spanning a range of masses;
 it also has the desirable feature that the function $\nu(y)$ converges
  very rapidly to 1 when $y \ga 20$, so deviations on Solar System 
  scales are predicted to be vanishingly small, consistent
  with observational limits.   

 For the case without the ExFE,  
  we apply Eq.~\ref{eq:g_mls} directly, with the conventional value 
  $a_0 = 1.2\times 10^{-10} \msecsq$.


To apply the External Field Effect (ExFE), we use the approximation 
 of \citet{BZ18}; this is given by 
\begin{eqnarray} 
 g_{N,int} & = & G(M_1 + M_2)/r^2  \\
 g_{N,gal} & = & 1.2 \, a_0 \\ 
 g_{N,tot} & = & \left( g_{N,int}^2 + g_{N,gal}^2 \right)^{1/2} \\ 
 g_{i, EFE} & = & g_{N,int} \nu(g_{N,tot}/a_0) \left(1 + 
    \frac{\kappa(g_{N,tot})}{3} \right) \\  
 \kappa & \equiv & \frac{\partial \ln \nu }{\partial \ln g_{N} } 
\label{eq:exfe} 
\end{eqnarray}
 where $g_{N,int}$ is the internal Newtonian acceleration of the binary; 
  $g_{N,gal}$ is the external (Galactic) Newtonian acceleration, 
 $g_{N,tot}$ is the quadrature sum of these, $\nu$ is the MLS function 
  from Equation~(\ref{eq:g_mls})  
  and $g_{i, EFE}$ is our model MOND-ian 
  internal acceleration, approximating the 
   application of the external field effect. 
 (This is not quite an exact solution of the MOND-like equations,
  but is shown by BZ18 to be a good approximation to the full 
   numerical solution).  

 Above,  the observed 
 Galactic rotation values 
  $v_{LSR} \simeq 232 \kms$ and $R_0 \simeq 8.1 \kpc$ imply 
 a total Galactic acceleration close to $1.75 \, a_0$, hence we 
  require $g_{N,gal} \nu(g_{N,gal}/a_0) \approx 1.75 \, a_0$. Solving this 
  leads to $g_{N,gal} \approx 1.16 \, a_0$ as above and 
  $\nu \approx 1.51$, in reasonably  
  good agreement with the estimated baryonic contribution to the Galactic
  rotation (as expected, since the MLS fitting function was derived by 
  fitting to a sample of external spiral galaxies with well-observed rotation
  curves, so this is consistent with our Galaxy being typical).  
 
 For the MLS $\nu$ function and ExFE approximation as above,  
 some example values for our wide-binaries 
  are  $g_{N,int} \sim 0.8 a_0, \ g_{N,gal} \sim 1.2 \, a_0$, so the
  internal and galactic accelerations are both comparable to $a_0$. 
  This leads to numerical values   
  $\nu(g_{N,tot} \sim 1.41 \, a_0) \sim 1.44$ 
  and $\kappa \approx -0.26$; this example leads to  
   $g_{i,EFE} \simeq 1.35 \, g_{N,int}$,  
  hence Eq.~\ref{eq:exfe} predicts that the wide binaries in ExFE 
  will  obtain a  $\sim 35\%$ boost  above Newtonian acceleration,  
   equivalent to a $\sim 16\%$ boost in typical binary orbital velocities. 
   This boost is significantly larger than we found in PS18 
   using the previous less accurate ExFE 
   approximation (Eq. 45 of PS18); hence this improves the prospects 
  for a decisive test (as was also briefly noted in PS18 Section~4.3).  

\begin{figure*} 
\begin{center} 
\includegraphics[width=17cm]{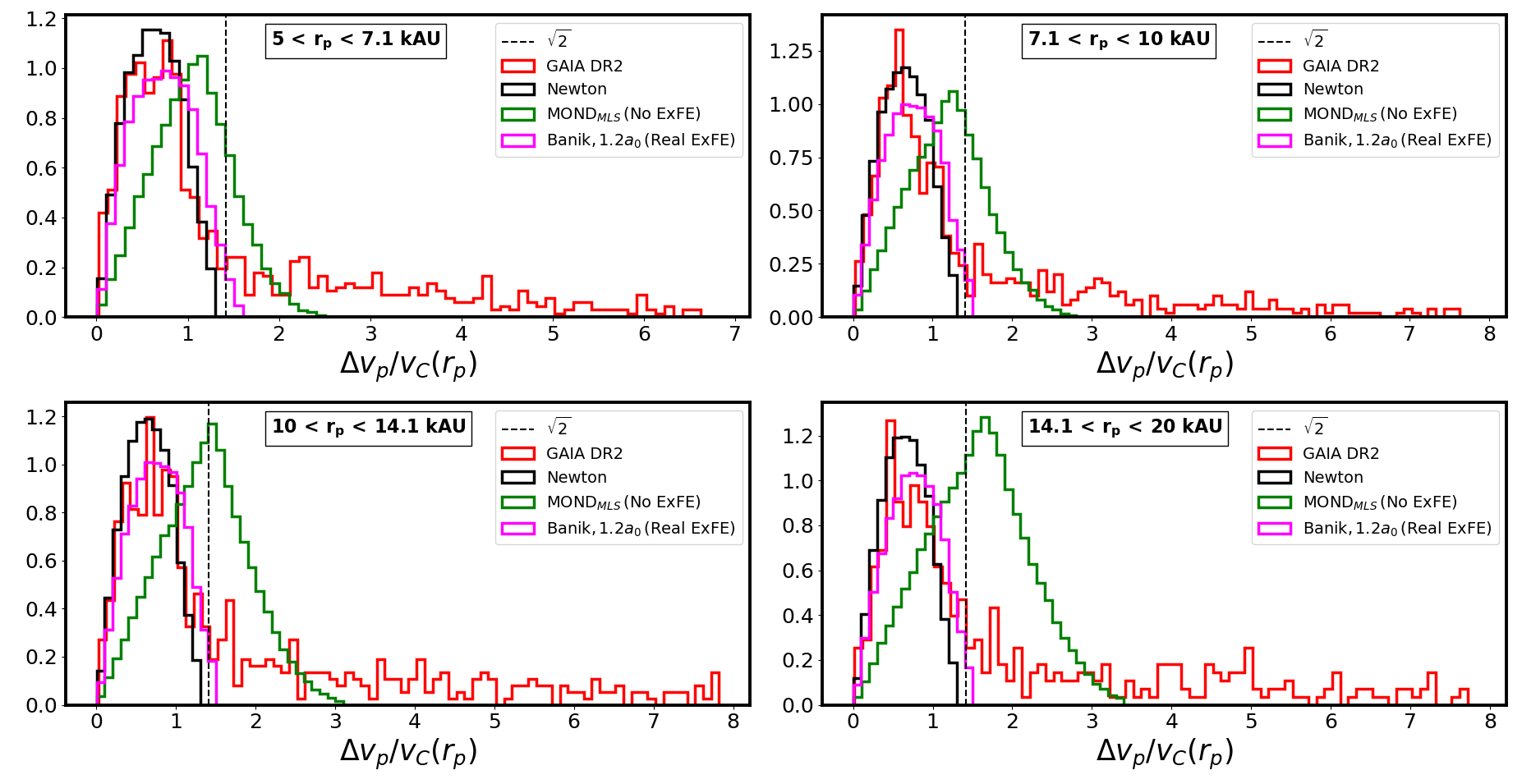} 
\caption{ 
  Histograms of velocity ratio, $\vratio$, for observed and simulated 
   binaries: the four panels show four bins of projected separation as 
  labelled in the legend.   The histograms show our observed 
   GAIA sample (red); 
  Newtonian simulated orbits (black); MOND {\em with} ExFE (magenta)
   and MOND {\em without} ExFE (green).   The simulated
  histograms are normalised to the number of observed systems
  at ratio $\le \sqrt{2}$.  } 
\label{fig:mghist} 
\end{center}
\end{figure*} 

 In Figure~\ref{fig:mghist}, we show histograms of 
 the ratio $\vratio$ for four bins of projected separation,
 and three gravity acceleration models: standard Newtonian,
  and MLS both without and with the ExFE, compared with the GAIA DR2 data. 
 Model histograms are normalised to match the total of the data at 
  velocity ratio $\le \sqrt{2}$. 

 On inspection of Figure~\ref{fig:mghist}, several features are
  immediately clear: 
\begin{enumerate} 
\item 
  The simulated histograms for MOND {\em without} ExFE show a large and obvious
  shift of the peak to larger velocity ratios, especially in the 
   wider separation bins where the Newtonian acceleration 
   is well below $a_0$;  
  this shifted peak appears clearly inconsistent with the data, 
    and plausible
   observational  errors or sample contamination 
   appear unlikely to remove this inconsistency. 
\item
 The simulated histogram for MOND {\em with} ExFE is much closer
  to the Newtonian case, but shows a distinctly larger fraction of binaries
  at velocity ratio $1.1 - 1.5$; the excess is increasing
  only moderately with projected separation, and saturates at still larger
  separations (beyond $20 \kau$).  
\item The actual data show a clear peak at $\sim 0.6$ as above, 
  then a fall towards $\sqrt{2}$; followed by 
   a prominent tail which slowly declines to much larger ratios $\ga 5$.  
   Clearly, the presence of this tail  makes it hard to decide a preference 
  between Newtonian or MOND-with-ExFE: a smooth downward extrapolation of
  the tail below $\sqrt{2}$  can  account for at least $50$ percent, possibly
   as many as 70 percent,  of the 
    observed number of systems with 
   velocity ratio $1.1 < \vratio < 1.4$, so 
   the tail is not well understood but
   has a major impact on the statistics.    
  This tail is discussed and modelled in the next Section. 
\end{enumerate}

\section{Flyby simulations} 
\label{sec:flyby} 

 We saw above that the observed histograms (Figure~\ref{fig:mghist}, 
   red lines)   
  of velocity ratios in GAIA DR2 can be qualitatively described 
  by a two-component ``hump + tail'' structure, 
  with the hump peaking near 0.6 in approximate agreement
 with Newtonian orbit expectations, while the smooth ``tail'' extends to much
 larger velocity ratios.  Our randomised-position samples show that
  the tail is much too populous to be explained by 
     random chance projections of unrelated stars;  
 one plausible explanation appears to be pairs of co-natal stars
  born in the same open cluster, which therefore have similar
  velocities and are currently undergoing a chance close flyby. 
 Evidence for a population of such ``cold streams'' has been
  given by \citet{Oh2017}. 

  In this case, we would expect two of the three velocity components
 (perpendicular to the escape direction) 
  must be similar in order to get a close flyby, while the velocity
 difference in the escape direction should approximately reflect
 the distribution in ejection velocities from the cluster
  and the time difference between the two ejections. 

 During a flyby, the relative velocity will speed up 
  according to a hyperbolic flyby orbit.  To simulate this, 
 we generated random flyby encounters as follows: 
 \begin{enumerate} 
\item We chose a distribution of impact parameter $b$ 
 with $dn/db \propto b$ up to a maximum value of $300 \kau$ or 1.45 pc. 
\item We chose various distributions of asymptotic velocity difference 
  $\vinf$, including a uniform distribution up to $2 \kms$ (hereafter the Flat
  distribution), two Maxwell-Boltzmann distributions with $\sigma = 0.75, 
  1.5 \kms$, and an exponential distribution 
  $\propto \exp(-\vinf / 1 \kms)$. 
\item Given values of $b, \vinf$ generated with distributions
  as above, each such pair produces one flyby encounter.  
   We sampled each hyperbolic flyby 
  at random times while the 3D separation was $\le 300 \kau$,
   and computed the 3D relative separation and relative velocity vectors. We 
  ``observed'' these from random viewing angles to produce $\vp, \rp$.  
\item We truncated the sample as for the DR2 data, 
  at $\rp < 40 \kau$ and  $\vp \le 3 \kms$, 
   and computed the resulting velocity ratio $\vp / v_c(\rp)$. 
\end{enumerate} 
  Clearly for hyperbolic flyby encounters the 3D velocity ratio $v_{3D}/v_c(r)$ 
  is always  $\ge \sqrt{2}$, with simulations showing a modest  
   pile-up in the distribution just above this value; this pile-up 
   arises from flybys with eccentricity
   not much larger than 1, which speed-up substantially but have
  velocity ratios decreasing towards $\sqrt{2}$ as 
  they approach pericenter.   
 The projection to 2D smears this distribution to lower ratios, 
   and thus fills-in the
   gap below $\sqrt{2}$; the result is that simulated fly-bys with a flat 
   distribution of $\vinf$ produce a smooth 
   maximum in the distribution at a ratio below 1.0, 
  with a gently declining  tail at larger ratios.      

\section{Data vs model comparisons}
\label{sec:comp} 
 
\begin{figure*} 
\begin{center} 
\includegraphics[width=7.5cm]{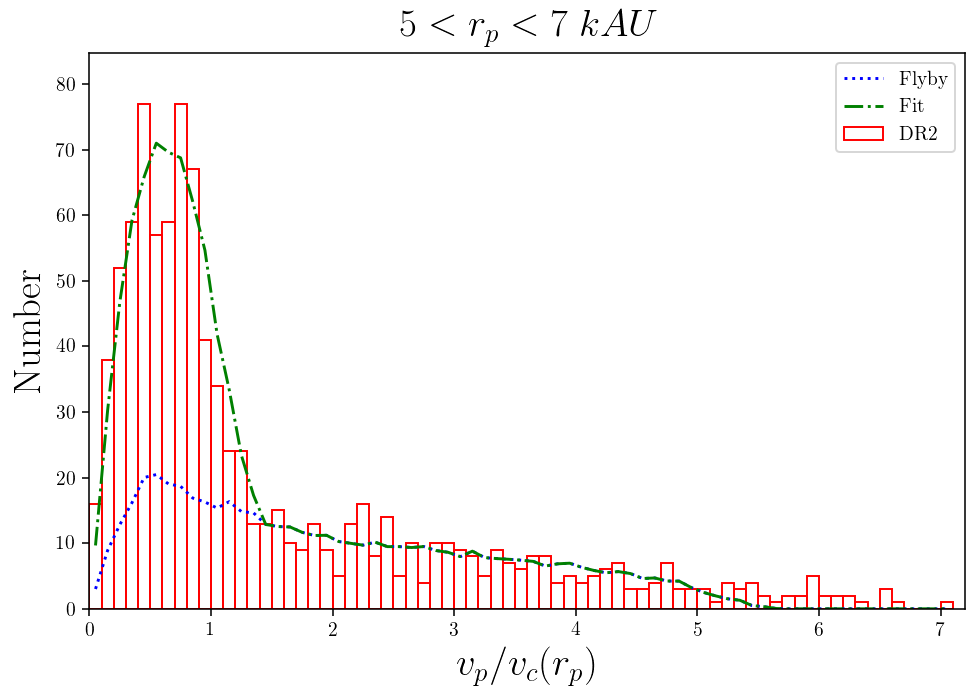} 
\includegraphics[width=7.5cm]{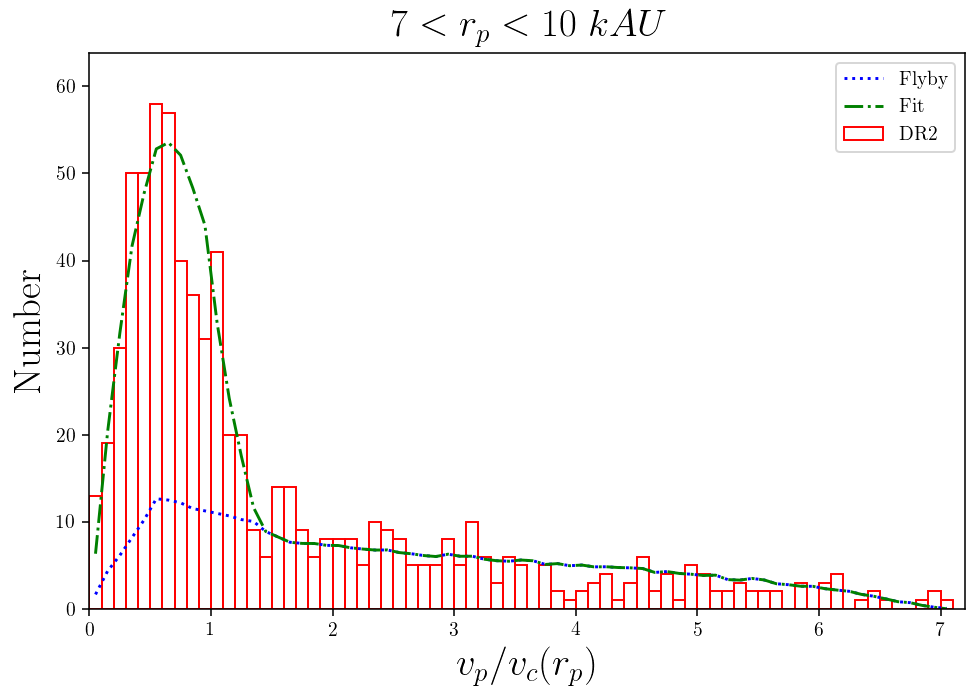} 

\includegraphics[width=7.5cm]{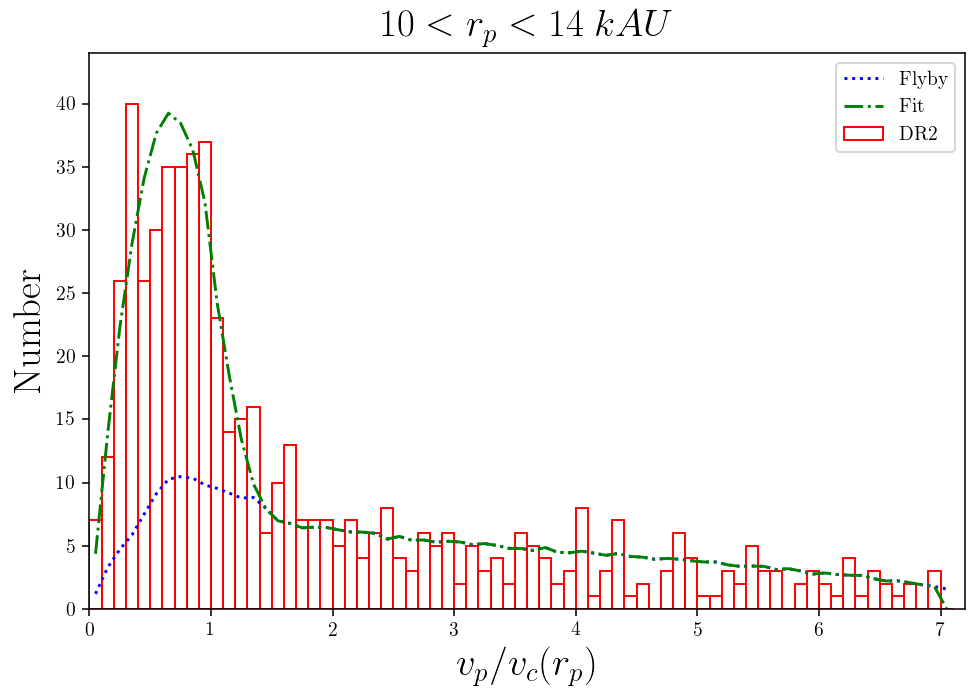} 
\includegraphics[width=7.5cm]{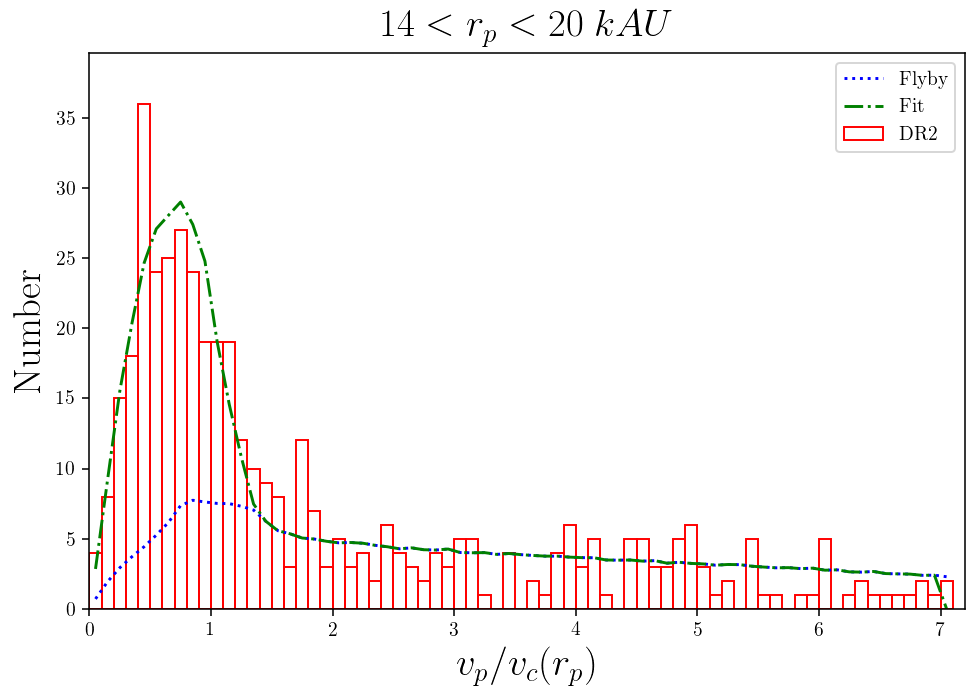} 

\caption{Fits of velocity ratios using a ``Binaries + Flybys'' model; 
  four panels show four bins of projected separation as labelled. 
  Here the Flyby model adopts a flat distribution of $\vinf$ up to 
  $2 \kms$;  fits adopt arbitrary normalisation for both binaries
 and flybys.  The red histogram shows the data. 
  The blue-dotted line is the fitted flyby population;
  green dashed line is the total, so the difference is fitted
   true binaries. } 
\label{fig:fits} 
\includegraphics[width=7.5cm]{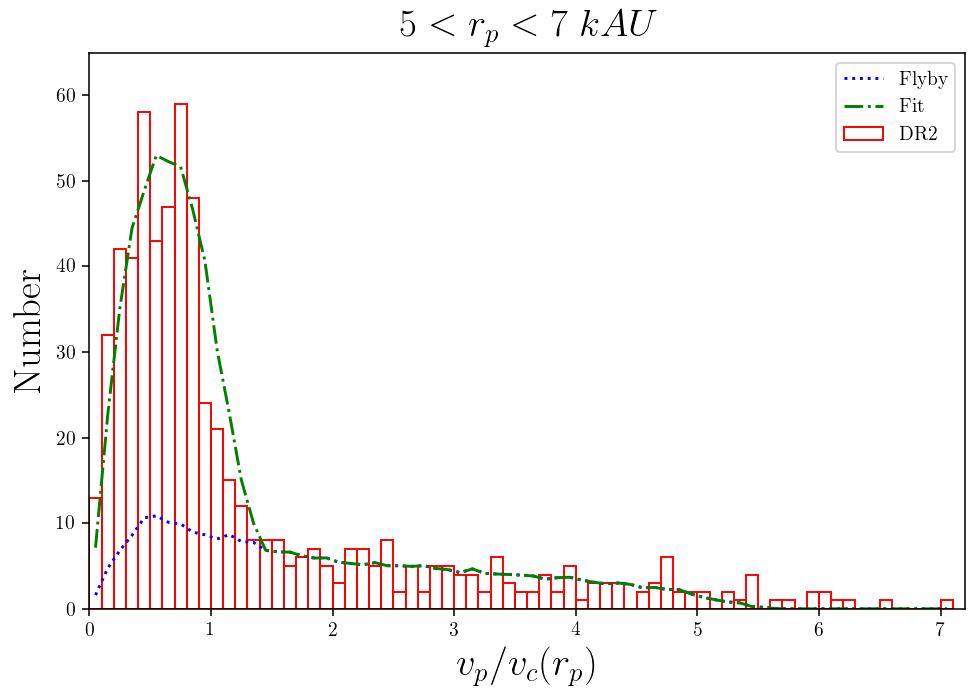} 
\includegraphics[width=7.5cm]{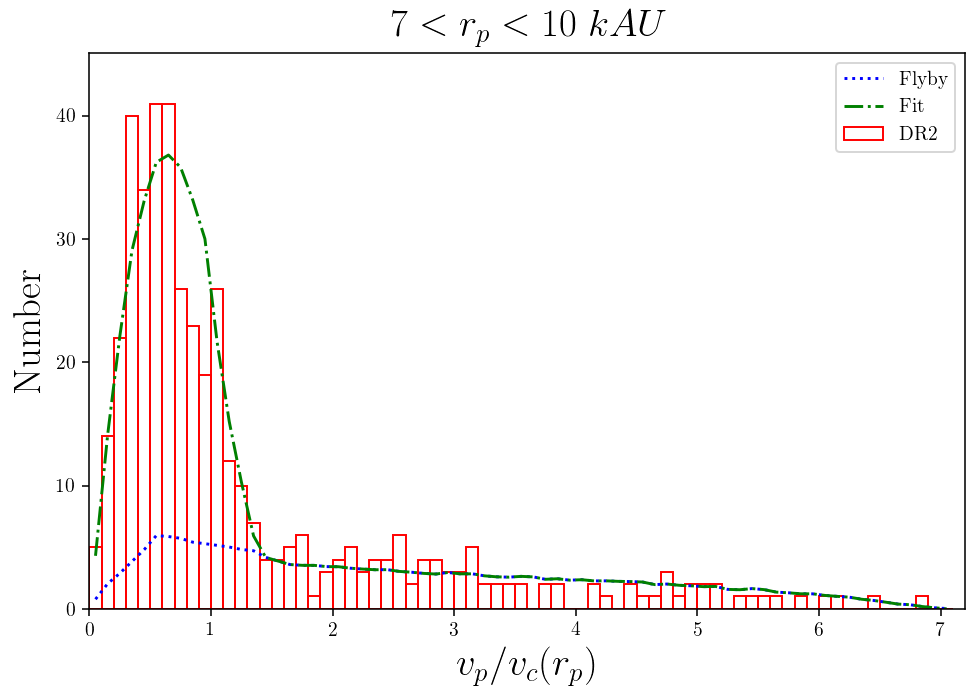} 

\includegraphics[width=7.5cm]{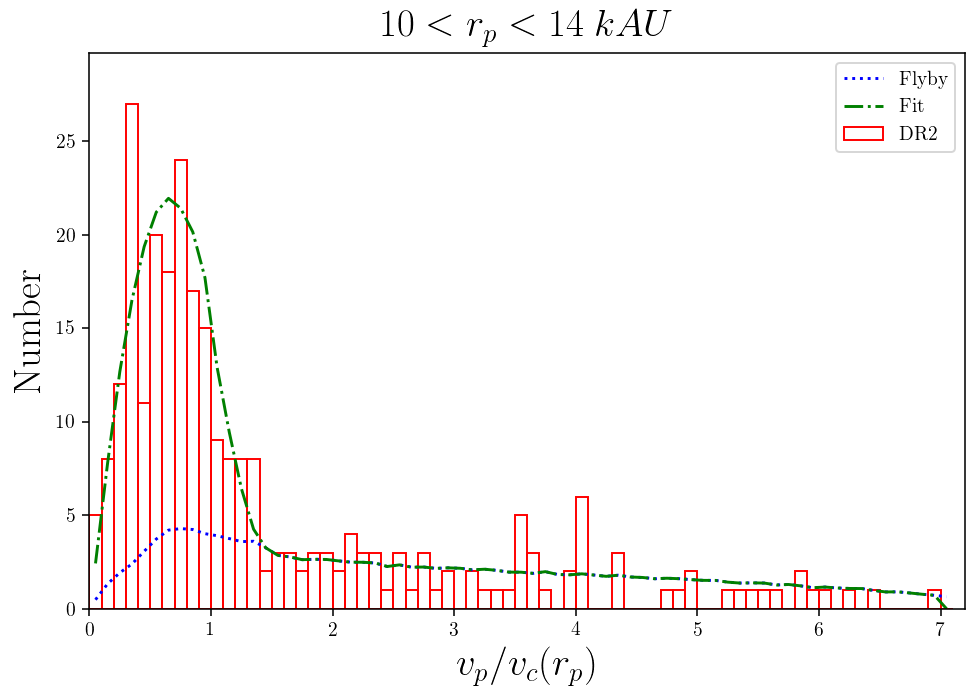} 
\includegraphics[width=7.5cm]{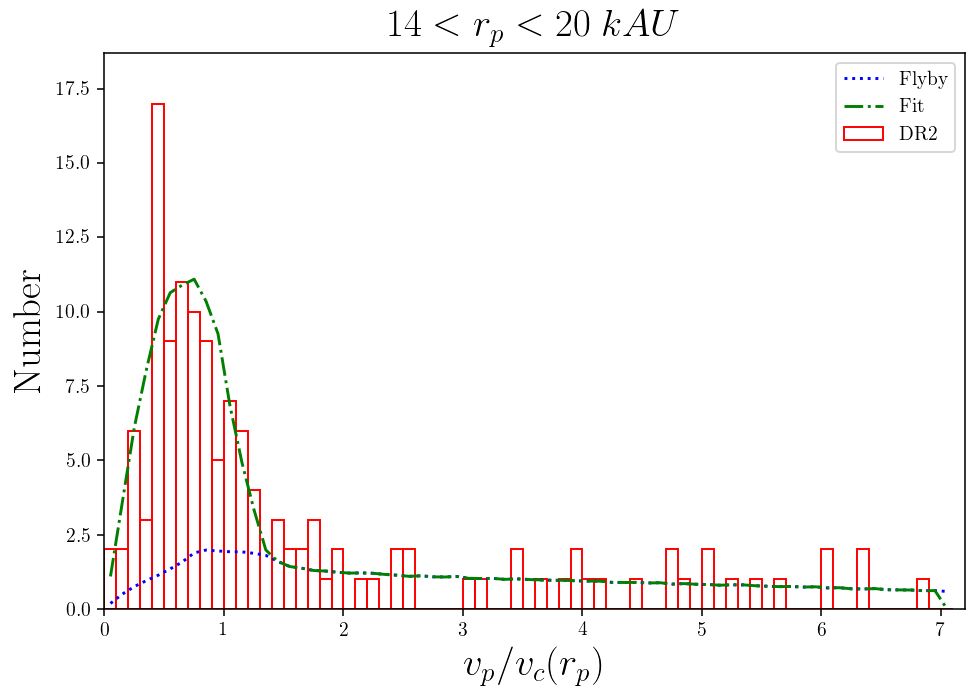} 
\caption{Same as Figure~\ref{fig:fits}, for the subsample of  
   binaries with $\sigma(\vp)/\vc(\rp) < 0.25$. } 
\label{fig:fits_e25} 
\end{center}
\end{figure*} 

 We now turn to a comparison between data and 
  models with a Binaries + Flybys population, 
 for the histograms of observed velocity
 ratio $\vp / v_c(\rp)$. We have produced histograms
 sliced in $\sqrt{2}$ bins of $\rp$, from $5 \kau$ to $20 \kau$; 
  hence four bins respectively 
  $5 - 7.1; 7.1-10; 10-14.1 ; 14.1-20 \kau$. These are chosen 
 spanning the most favourable range for the gravity test, 
  since in PS18 it was found that modified-gravity effects 
  with the ExFE included should largely saturate above $10 \kau$; 
  while the statistics
  become rather poor and contamination worsens beyond $\rp > 20 \kau$.
 More recent simulations from \citet{BZ18} (which used a more realistic
   formulation of the ExFE compared to PS18) 
  led to a similar conclusion on this optimal separation range.   

For fitting, we take the simulated Newtonian binaries with
 the Tokovinin eccentricity distribution, and the simulated
  hyperbolic flybys with the Flat $\vinf$ distribution. 
 We produce a 2-parameter fit by keeping the shapes of the Binary
  and Flyby populations fixed,  and simply adjusting 
  the relative normalisations of
    each of the Binary and Flyby populations, fitting the sum of these  
  to the observed histogram assuming Poisson errors 
  to provide a maximum-likelihood  fit.   Results of this fit 
  are shown in   Figure~\ref{fig:fits},  and it is seen that the fits 
   match the data quite well given the statistical noise; this is
  confirmed by the $\chi^2$ values which are acceptable.  
  Counts of the observed and fitted histograms in some selected
  ranges of velocity ratio are given in Table~\ref{tab:stats}.

\begin{table*}
\caption{Number of candidate binaries in selected ranges of 
  projected separation and velocity ratio: data, and model fits for
  combined Newtonian (N) and flyby (F) populations. 
  Newtonian+flyby fits. Rows are range of projected separation, as
  in Column 1.    Columns 2-4 are for all velocity ratios, 
   columns 5-7 for ratios $< 1.4$, and columns 8-10 for
    ratios between $1.1$ and $1.4$.  }
\label{tab:stats}  
\centering 
\begin{tabular}{l|rrr|rrr|rrr|} 
\phantom{      } &  \multicolumn{3}{|c|}{$\vratio < 7$} & 
   \multicolumn{3}{|c|}{$\vratio < 1.4$} & 
 \multicolumn{3}{|c|}{$1.1 < \vratio < 1.4$} \\ 
\hline 
 $\rp$ range  &   Data & Fit(N)  & Fit(F)  & Data & Fit(N) & Fit(F) 
        & Data & Fit(N)   & Fit(F) \\ 
\hline 
 $5 - 7.1 \kau$  & 955 & 440.5 & 513.4 &  638 & 440.5 & 213.5 & 61 & 28.3 & 45.8 \\
 $7.1 - 10 \kau$  & 711 & 352.3 & 390.4 & 474 & 352.3 & 132.5 & 49 & 22.4 & 30.9 \\
 $10 - 14.1 \kau$  & 569 & 244.6 & 353.4 & 352 & 244.6 & 108.8 & 45 & 15.0 &  26.8 \\
 $14.1 - 20 \kau$  & 434 & 183.6 & 282.5 & 260 & 183.6 & 77.2 & 41 & 11.0 & 21.8 \\
\hline 
\end{tabular}
\caption{Same as Table~\ref{tab:stats}, but for the subsample of binaries
   with relative velocity error ratio $\sigma(\vp)/\vc(\rp) < 0.25$. } 
\label{tab:stats_e25} 
\begin{tabular}{l|rrr|rrr|rrr|} 
\phantom{      } &  \multicolumn{3}{|c|}{$\vratio < 7$} & 
   \multicolumn{3}{|c|}{$\vratio < 1.4$} & 
 \multicolumn{3}{|c|}{$1.1 < \vratio < 1.4$} \\ 
\hline 
 $\rp$ range  &   Data & Fit(N)  & Fit(F)  & Data & Fit(N) & Fit(F) 
        & Data & Fit(N)   & Fit(F) \\ 
\hline 
 $5 - 7.1 \kau$  & 629 & 367.0 & 272.5 &  463 & 367.0 & 113.3 & 35 & 23.5 & 24.3 \\
 $7.1 - 10 \kau$  & 428 & 265.7 & 183.5 & 320 & 265.7 & 62.2 & 29 & 16.9 & 14.5 \\
 $10 - 14.1 \kau$  & 270 & 149.7 & 144.8 & 190 & 149.7 & 44.5 & 24 & 9.2 &  11.0 \\
 $14.1 - 20 \kau$  & 134 & 78.2 & 72.2 & 91 & 78.2 & 19.7 & 10 & 4.7 & 5.6 \\
\hline 
\end{tabular}
\end{table*}

The fits indicate that over the interval $0 \le \vp / v_c(r_p) \le \sqrt{2}$, 
   the Binary population dominates over the model Flybys by a factor of 
   2 - 4, i.e. most but not all such systems are genuine binaries.  

 However, it is also seen from the fits that 
  the situation is reversed in the most interesting 
   range $1.1 - 1.4$; here the model Flyby population 
  dominates over the model Binary population by a factor of
 $\sim 1.5 - 2.5$.  While
 this is not necessarily fatal since the flyby population 
  can be statistically subtracted, it does imply that
  a better physical understanding or modelling of the flyby population 
  will be required to derive conclusions about MOND with ExFE.  
 We postpone this modelling to a future paper. 

 Concerning the effect of proper-motion errors on the velocity
  histograms, we have repeated the above procedure for the subsample
 of binary candidates with an additional cut $\sigma(\vp)/\vc(\rp) < 0.25$; 
  the corresponding results are shown in Figure~\ref{fig:fits_e25} 
  and Table~\ref{tab:stats_e25}. This does reduce the sample 
  size quite substantially,  by a factor $\sim 2/3$ in the first bin
   down to $\sim 1/3$ in the widest bin;  however, the tail is
 still present and the general appearance of the histograms is
  approximately unchanged.  
 The fitted ratios of Flyby vs Newtonian binaries in the 
  range of velocity ratio 1.1 -- 1.4 do slightly
   reduce, but this is countered by the small-number statistics so
   again no firm conclusion concerning MOND with ExFE can be made
   at present. 
  However, with the substantial reduction of
  proper motion errors expected in the future GAIA DR3, 
  a large majority of candidate binaries 
  will be expected to survive the above cut.   

 We also note that while the flyby model does provide an acceptable
 fit to each of the individual histograms,  there is a potential
  inconsistency in that the fitted number of ``flyby'' events
 is slightly decreasing with $\rp$, while the simulations predict
 a rising distribution.  This suggests that there may be an
 additional contribution to the tail from e.g. undetected hierarchical
 triples or non-Gaussian errors in the GAIA proper motions. 
  Further work e.g. with radial velocities and future GAIA
  data releases is probably required to understand the origin of the tail.

\subsection{Discussion and future prospects} 

We have seen that the observed distributions of velocity
 ratio for our candidate binaries appears to be fairly strongly inconsistent
 with MOND {\em without} ExFE, since the observed peak stays close to
  the Newtonian prediction $\sim 0.6$ independent of $\rp$, 
   and there are many more observed 
 systems  with $\vratio < 1$ compared to $1 < \vratio < 2$, 
  contrary to the model values of MOND without ExFE. 
  Realistic contamination or observational errors is expected 
  to produce more contaminants at $1 < \vratio < 2$ than $\vratio < 1$, 
  hence is unlikely to erase this discrepancy.   

 We also find that the number of binary candidates 
 is sufficient {\em in principle} for the more challenging
 case of testing MOND {\em with} ExFE, e.g. we have a total
 of 1724 binary candidates with $5 < \rp < 20 \kau$ and 
  $\vp / \vc(\rp) < \sqrt{2}$;  given this, Newtonian
  models predict $\la 95$ above ratio $1.1$ while MOND-ExFE predicts
  $\ga 200$, values which would be easily separable at high significance 
  if simple Poisson statistics applied. 

 But, the presence of the high-velocity tail is currently 
 poorly understood; the tail is much too populous to be explained
 by chance coincidences, but extends to high velocity ratios. 
 We note that the presence of this tail will severely
  contaminate any statistic based mainly on {\em rms} velocity differences; 
 thus, previous hints of excess {\newtwo binary} {\em rms} velocities 
   in the literature (e.g. \citealt{Hernandez 2019})  may well
  be caused by this tail, rather than an actual modification of gravity. 
 {\newtwo It therefore appears that improved modelling or understanding
  of the origins of this tail will be crucial in future when 
 trying to test Newtonian gravity against MOND-ExFE type modifications; 
  in particular, it will be necessary to quantify the contribution of 
  non-bound flyby systems or hierarchical triples to the important 
 velocity window $\vratio \sim 1.1 - 1.4$, which may be challenging. }  

{\newtwo Hierarchical triples with an undetected third object are another
 possible contributor to the tail, though the tail seems too populous
  for hierarchical triples to contribute a majority of the tail.  
  Most such systems can be detected in principle by either radial
 velocity variation over the years (for close-in third stars), or detection 
  in high-resolution imaging (for wider third stars).} 
    
 The prospects for improved data in the future are good:
 the anticipated GAIA DR3 should provide a factor-2 improvement
 in proper-motion precision, along with many more epochs to weed out
 anomalous non-Gaussian errors; the extended mission to 2022+ should
  provide another factor-2. This will also allow an expanded
 sample, e.g. pushing the distance limit moderately outward
   to $\sim 250 - 300 \pc$.  
 Spectroscopic observations can deliver the missing radial 
 velocity information; this improves the statistics, by increasing
 the predicted fraction of bound binaries 
   in the key range $\vratio \sim 1.1 - 1.4$; 
 {\newtwo while} most (but not all) unbound flyby systems should move 
  to ratios above 1.5 with radial velocities included, so this will help 
   to sharpen the discrimination. 

 Also, modelling of ``cold streams'' may be helpful: it is currently
 difficult to detect poor cold streams due to angle-dependent 
   projection effects in 2D data, 
 but if modest-precision radial velocities ($\sigma \sim 2 \kms$)
  become available for a good fraction of 
  stars with $G \la 16$, this would
 allow much cleaner selection of cold streams via matching 3D velocities; 
  we could then reject ``binaries'' consistent  
  with membership in streams. 
 (We note that much higher RV precision $\sim 0.1 \kms$ is required to
 get 3D velocities for surviving wide binary candidates,
  but rejection of most cold streams should be possible 
  with only $2 \kms$ RV precision).  

 Selecting stars by age, if possible, should also be helpful: most binaries at
  $10 \kau$ should survive for the age of the Galaxy, while common-origin
 unbound pairs or cold streams should disperse over Gyr timescales
  and thus show a strong  bias to young ages. Thus, if it were possible
  to add an additional selection cut on age $\ga 3$ Gyr,
    we could substantially reduce
  contamination from unbound co-natal flyby systems.  

 To summarise, we have seen that the {\newtwo observed } 
  population of wide binaries
 in GAIA DR2 already disfavours MOND {\em without} ExFE. 
 For the more interesting case of MOND {\em with} ExFE,  
  there are likely an adequate 
  number of probable bound binaries to carry out the test as outlined 
   in PS18;  
  the current velocity precision in GDR2 is slightly marginal, 
   but the future GAIA DR3/DR4 should be readily precise enough, 
  that observational followup data on candidate binaries obtainable 
  within a 5-7 year timescale 
  has good prospects for actually supporting or refuting 
  acceleration-based models of modified gravity  
   similar to MOND-with-ExFE, subject to a better understanding of the
   origins of the high-velocity tail.  



\section{Conclusions}
\label{sec:conc} 

We have used the recent GAIA Data Release 2 to select 
 a large sample of candidate wide binary stars at $d \la 200 \pc$
 and magnitude $G < 16$, which are suitable 
  for testing modifications of gravity at
  $g \la a_0$.  We applied various cuts to minimise contamination, 
 leading to a cleaned sample of $24,282$ candidate binaries,
  with $2,749$ in the wide-separation range $5 - 20 \kau$.  
 After estimating masses from a main-sequence mass/luminosity relation,
  we derived for each candidate binary the ratio $\vp / v_C(\rp)$. 
 We compared this sample to various control samples with randomised
 positions; we find that the real sample is much more numerous, hence
   chance projection systems are relatively negligible.  
  We then cut the samples into $\sqrt{2}$ bins of $\rp$, 
  and explored the histograms of velocity ratio for each separation bin, 
  and made various comparisons with models for Newtonian and MOND-like
  gravity models, and contamination from flyby populations.  

Our main conclusions are as follows: 
\begin{enumerate} 
\item The number of candidate binaries in the real data is much 
 larger than in the randomised control samples, by a factor 
 $\sim 100$ at $5 \le \rp \le 7.1 \kau$, down to $\sim 11$ in the
 bin from $14 - 20 \kau$.  
  Thus, the large majority of our candidate binaries are clearly 
 physically associated in some way, 
  in agreement with \citet{Andrews 2017} and
 \citet{Andrews 2018}.   The frequency of ``probable bound'' binaries
 is declining beyond $5 \kau$, approximately in agreement with 
  \citet{Andrews 2018}.   

\item The histograms of relative velocity 
  are well described by a quasi-Newtonian ``peak''
  (presumably bound binaries), plus a gently declining ``tail'' which extends
  to substantially larger velocity ratios $\ga 5$.  The tail extends
  to velocity ratios larger than any reasonable modified-gravity model;
  we speculate that the tail is likely to 
 result from co-natal star pairs originating in the same open cluster, 
  which are unbound but currently undergoing a close flyby 
  at a relative velocity  $\sim 1 - 3 \kms$. 

\item The existence of the high-velocity tail implies that previous hints of
  anomalous velocities based on {\em rms} statistics are likely
   contaminated. 

\item MOND-like theories {\em without} an ExFE predict a substantial
  shift of the peak at large separations $\sim 10 - 20 \kau$, which is
 not observed; these models appear to be strongly disfavoured by the
  current data. 

\item 
  Precision tests of MOND-like theories {\em with}
 the ExFE are not quite practical giving the current data,
 but should be possible in the near future,  given improved modelling 
  and understanding of the high-velocity tail.  
 Further data such as high-precision radial velocities,
 removal of cold streams and age estimates should be helpful 
  in this direction.  

\item The available data should improve quite rapidly in the near future:
  in particular, the upcoming GAIA DR3 will provide a substantial
 improvement in proper motion precision, 
  allowing a larger usable distance limit and a larger sample.  
  Future large-multiplex $R \sim 20,000$ spectrographs such as
 4MOST, WEAVE and MSE should allow high-quality spectra to be 
 obtained for most of these candidate binaries.  Thus with 
  anticipated data over the next $\sim 5$ years, it looks promising
 that wide binaries can provide an interesting and 
 direct observational test for  possible modifications of gravity 
   at low accelerations $\sim 10^{-10} \msecsq$.  

\end{enumerate}

\section*{Acknowledgements}
We thank Indranil Banik and Tim Clifton for helpful discussions. 
 CP has been supported by an STFC studentship.  

This is an author-produced, non-copy-edited version of the paper as
accepted by MNRAS. The version of record is available at 
 DOI:10.1093/mnras/stz1898  . 








\end{document}